\documentclass[a4paper,11pt,3p,onecolumn]{quantumarticle}
\pdfoutput=1
\usepackage{graphicx}
\usepackage{amsmath,amsthm,amssymb,mathtools,mathrsfs}
\usepackage{bm}
\usepackage{subcaption}
\usepackage{ulem}
\usepackage{algorithm}
\usepackage{algorithmic}
\usepackage{multirow}
\usepackage{adjustbox}
\usepackage[table,xcdraw]{xcolor}
\usepackage{hyperref}
\usepackage{tabularx}
\usepackage[squaren,Gray]{SIunits}
\usepackage{xcolor}
\usepackage{soul}
\usepackage{nicefrac}
\usepackage{float}
\usepackage{pdfcomment}
\usepackage{eqparbox}
\usepackage{lineno}
\usepackage{nicematrix}
\usepackage{setspace}
\usepackage{empheq}
\usepackage[numbers]{natbib}

\usepackage[english]{babel}

\begin{document}

\title{Quantum algorithm for bioinformatics to compute the similarity between proteins}

\author[expleo]{Anthony Chagneau}
\email{anthony.chagneau@expleogroup.com}
\author[expleolyon]{Yousra Massaoudi}
\author[expleolyon]{Imene Derbali}
\author[expleolyon]{Linda Yahiaoui}
\affil[expleo]{Expleo Group, Agence M\'{e}diterran\'{e}e, 2 Impasse de Chasles, Z.A Cap Horizon, Vitrolles, 13127, France}
\affil[expleolyon]{Expleo Group, 21 Rue Andr\'{e} Lwoff, Saint-Priest, 69800, France}

\begin{abstract}
Drug discovery has become a main challenge in our society, following the Covid-19 pandemic. Even pharmaceutical companies are already using computing to accelerate drug discovery. They are increasingly interested in Quantum Computing with a view to improve the speed of research and development process for new drugs. Here, the authors propose a quantum method to generate random sequences based on the occurrence in a protein database and another quantum process to compute a similarity rate between proteins. The aim is to find proteins that are closest to the generated protein and to have an ordering of these proteins. First, the authors will present the construction of a quantum generator of proteins who define a protein, called the test protein. The aim is to have a randomly defined amino-acids sequence according to a proteins database given. The authors will then describe two different methods to compute the similarity's rate between the test protein and each protein of the database and present results obtained for the test protein and for a case study, the elafin.
\end{abstract}

%========================================================================================
\section{Introduction}
%========================================================================================
Nowadays, following the Covid-19 pandemic, it is necessary to accelerate drug discovery. Computational approaches are widely used because drug discovery is a very complex, time consuming and resource intensive process. These approaches have attracted considerable interest due to their potential to accelerate drug discovery in terms of time and cost. The development of a new drug usually takes more than ten years and about a billion dollar budget \citep{McKinsey2021, Accenture}. In drug discovery early phase, computational methods are widely applied to decipher disease-related biology, prioritise drug targets and identify and optimise new chemical entities for therapeutic intervention. In general, the main objectives of in silico approaches in drug discovery include the generation of better compounds with desirable in vitro and in vivo properties. In addition, the computational analysis essentially assists in the orientation of experimental programmes by selecting the number of candidate compounds to be experimentally evaluated. Many drug compounds have been successfully developed using computational methods, but in reality more than 50\% of drug discovery candidates fail in early phase clinical trials \citep{McKinsey2021}.

The identification and development of small molecules and macromolecules, that can help cure diseases, is the core business of pharmaceutical companies. The pharmaceutical industry is a natural candidate for Quantum Computing (QC) given its focus on molecular formations. Molecules, including those that could be used as drugs, are in fact quantum systems, i.e., systems based on the quantum physics phenomena. 

\begin{figure}[H]
\centering
\includegraphics[width=0.99\textwidth]{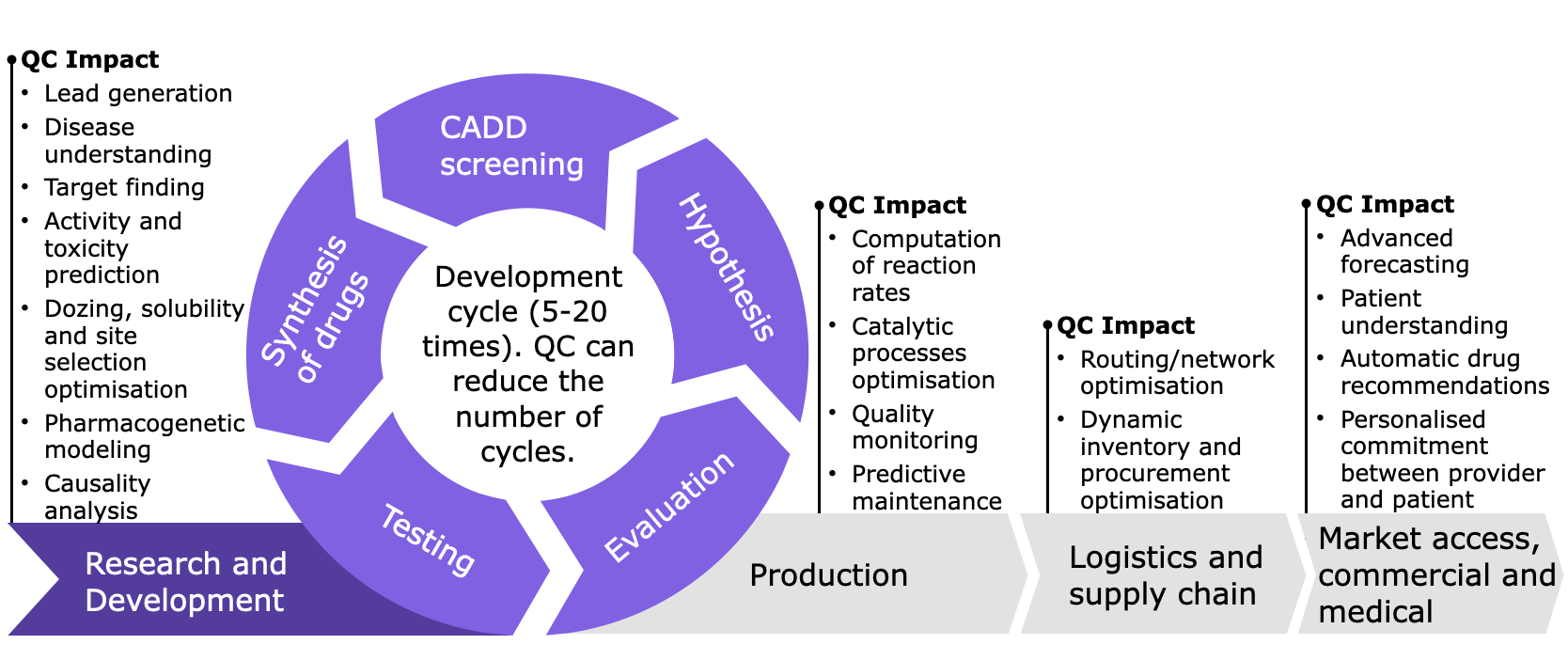}
\caption{Effects of quantum computing (QC) on the pharmaceutical value chain}
\label{McKinsey}
\end{figure}

Currently, the pharmaceutical industry processes molecules using non-QC tools in a methodology known as computer-aided drug discovery (CADD). But the classical computers are severely limited and basic calculations like predicting the behaviour of medium-sized drug molecules can take a great deal of time to compute accurately. CADD on quantum computers could improve the scope of biological mechanisms suitable for CADD. And it could also reduce the length of the empirical development cycle by eliminating some of the research \textquotedblleft{}dead ends\textquotedblright{} that add a lot of time and money to the discovery phase \citep{McKinsey2021}. The impacts of quantum computing on the pharmaceutical value chain are represented in Figure \ref{McKinsey}.

The main interest of quantum algorithms in the field of drug discovery is to decrease the cost and time of the research and development process of new drugs as it is true in other fields. Quantum computing is being explored in other fields, such as communication \citep{Breiner19,Chakraborty23}, physics simulations \citep{Childs21,Zanger21,Shokrian23,Chagneau23}, machine learning \citep{Mari20,Roggero22,Landman22,Marshall23} and finance \citep{An21,Zhuang23}. Today, some quantum methods have already been developed to represent molecules systems \citep{Kottmann23}. Thus, when fully developed, QC could add value to the entire drug value chain, from discovery through development to registration and post-marketing.

The aim of this paper is to use quantum computing to design a random amino-acids sequence of a protein, called the test protein, and to search for similar proteins to the test protein in a database.  All using algorithms with a quantum formalism.

%========================================================================================
\section{Material and methods}
%========================================================================================
In our study, two main main lines of work are identified: the construction of the structure of a protein by its amino-acids sequence and the computation of a similarity's rate between proteins.

\begin{figure}[H]
\centering
\includegraphics[width=0.99\textwidth]{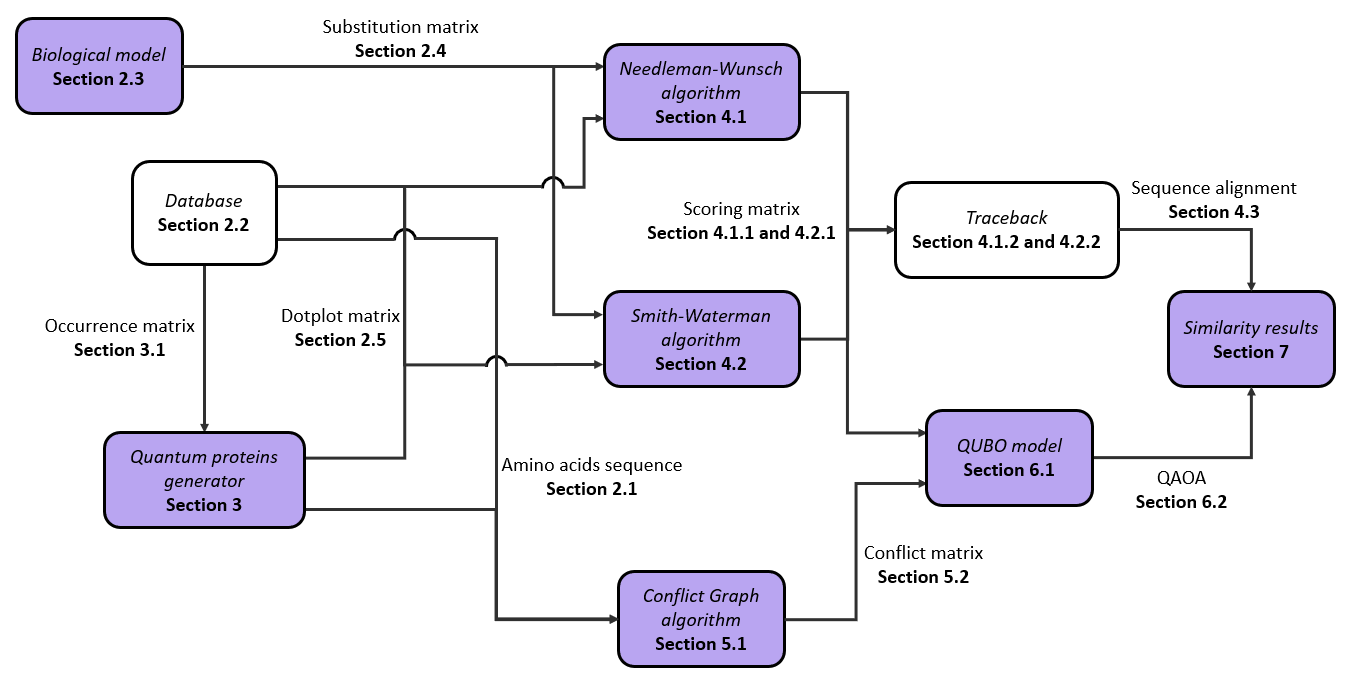}
\caption{Process framework}
\label{Framework}
\end{figure}

The first idea here is to construct an amino-acids sequence with a quantum algorithm and more precisely with the measurement of a quantum system. The second idea is to compute the similarity's rate between two amino-acids sequences. Here, we build three quantum algorithms with different methods and compare them afterwards. A framework describing the process is presented in Figure \ref{Framework}. 

%========================================================================================
\subsection{Amino-acids sequence}
%========================================================================================
Amino acids are the component that build proteins and thus provide their structure \citep{Wu09}. The distinction between the different amino acids is made by more or less complex side chain which gives them different physico-chemical properties \citep{Neilan03}. Amino acids can be linked together by a peptide bond to form chains containing from two to several thousand amino acids. This chain, called polypeptide chain, is the ordering of the amino acids in the sequence. The protein sequences present in a living organism are encoded by DNA, and more specifically by its nucleotides \citep{Yachie09}.

The structure of a protein is determined by the sequence of amino acids that make it up and how the protein folds into more complex shapes \citep{Scheeff03}:

\begin{itemize}
\item Primary structure: amino-acids sequence
\item Secondary structure: local interactions between the different amino acids.
\item  Tertiary structure: three-dimensional structure of the protein.
\item Quaternary structure: interactions of different proteins to form more complex proteins.
\end{itemize}

In this article, we will focus only on the primary structure with the twenty $\alpha$-amino acids, called also the standard proteinogenic amino acid, that make up proteins. This different amino acids are presented in the Table \ref{Aminoacid}.

\begin{table}[h!]
\centering
\resizebox{\textwidth}{!}{%
\begin{tabular}{|l|c|c|c|}
\hline
\rowcolor[HTML]{9698ED}  
\textbf{Name} & \textbf{Abbreviation} & \textbf{Symbol} \\ \hline 
\rowcolor[HTML]{EBE9F5} 
Alanine & Ala & A   \\ \hline 
\rowcolor[HTML]{EBE9F5} 
Arginine & Arg & R  \\ \hline 
\rowcolor[HTML]{EBE9F5} 
Asparagine & Asn & N  \\ \hline 
\rowcolor[HTML]{EBE9F5} 
Aspartic acid & Asp & D  \\ \hline 
\rowcolor[HTML]{EBE9F5} 
Cysteine & Cys & C  \\ \hline 
\rowcolor[HTML]{EBE9F5} 
Glutamine & Gln & Q  \\ \hline 
\rowcolor[HTML]{EBE9F5} 
Glutamic acid & Glu & E  \\ \hline
\rowcolor[HTML]{EBE9F5}  
Glycine & Gly & G  \\ \hline
\rowcolor[HTML]{EBE9F5}  
Histidine & His & H  \\ \hline
\rowcolor[HTML]{EBE9F5}  
Isoleucine & Ile & I   \\ \hline 
\end{tabular}
\begin{tabular}{|l|c|c|c|}
\hline
\rowcolor[HTML]{9698ED}  
\textbf{Name} & \textbf{Abbreviation} & \textbf{Symbol} \\ \hline 
\rowcolor[HTML]{EBE9F5} 
Leucine & Leu & L  \\ \hline
\rowcolor[HTML]{EBE9F5}  
Lysine & Lys & K  \\ \hline
\rowcolor[HTML]{EBE9F5}  
Methionine & Met & M   \\ \hline
\rowcolor[HTML]{EBE9F5}  
Phenylalanine & Phe & F   \\ \hline
\rowcolor[HTML]{EBE9F5}  
Proline & Pro & P  \\ \hline
\rowcolor[HTML]{EBE9F5}  
Serine & Ser & S   \\ \hline
\rowcolor[HTML]{EBE9F5}  
Threonine & Thr & T   \\ \hline 
\rowcolor[HTML]{EBE9F5} 
Tryptophan & Trp & W   \\ \hline 
\rowcolor[HTML]{EBE9F5} 
Tyrosine & Tyr & Y  \\ \hline
\rowcolor[HTML]{EBE9F5}  
Valine & Val & V  \\ \hline
\end{tabular}
}
\caption{List of amino acids}
\label{Aminoacid}
\end{table}

These amino acids serve both as a protein quantum generator, since they build an amino-acids sequence, and as the first elements to be compared when looking for similarities between two proteins. 

%========================================================================================
\subsection{Databases and BLASTP}
%========================================================================================
In order to design therapies, it is first necessary to search for and select relevant, consistent and interpretable data. To do this, we began by searching and selecting various databases containing a large amount of information for the pharmaceutical industry. In total, we selected six databases on the basis of their use and content:

\begin{itemize}
\item HMP: Genomics data,
\item Amadis: Genomics data,
\item Disbiome: Genomics data,
\item VMH: Genomics and Metabolomics data,
\item MicrobiomeDB: Genomics, Transcriptomics and Proteomics data,
\item Uniprot: Proteomics data.
\end{itemize}  

In the first instance, we are only interested in proteins and only in their structure (amino-acids sequence), we are restricting ourselves to Uniprot. We consider the reviewed version of Uniprot, i.e., Swiss-Prot with 568 744 proteins at the time the article was written. Documents are available in different formats such as JSON, GFF, RDF, SMILES and FASTA. For our study, we will consider the FASTA format. In the FASTA format file representing the Uniprot database, the following information is provided for each protein:

\begin{itemize}
\item Number of the UniProt entry,
\item Identifier of the UniProt entry,
\item Protein name of the UniProt entry,
\item Scientific name of the organism of the UniProt entry,
\item Unique identifier of the source organism,
\item Numerical value describing the proof of the protein's existence,
\item Sequence version number,
\item Sequence of amino acids.
\end{itemize}

We need a reference to compare our results. This will be the BLASTP tool (Protein Basic Local Alignment Search Tool), which is based on a heuristic search method. BLASTP finds regions of similarity between biological sequences. The tool compares one or more protein sequences with amino-acids sequence databases and calculates the significance threshold, i.e., the threshold at which the results of the comparison are deemed reliable. BLASTP works in three stages \citep{Dardel2002}:

\begin{itemize}
\item Decomposes the test sequence into overlapping segments of a given length and search for all possible segments with a homology score above a given threshold. BLASTP then builds a dictionary of all segments with minimum local homology;
\item Scans the database using the previously created dictionary. Whenever BLAST identifies a match with an amino-acids sequence in the database, it attempts to extend this homology upstream and downstream of the segment initially found;
\item Once the homology has been extended, it uses a scoring system to estimate the probability that the match is due to a random occurrence.
\end{itemize}

%========================================================================================
\subsection{Biological model}
%========================================================================================
In the course of evolution, from one generation of an individual to the next, protein sequences can be progressively modified by DNA mutations. Each amino acid is more or less likely to mutate into another amino acid \citep{Jayaraman22}. These mutations are mainly due to the repetition of the genetic code, which translates similar codons (groups of three nucleotides) into the appropriate amino acids. These amino acid changes occur at the level of DNA structure, affecting one or more nucleotides (between one and ten). There are four types of nucleotide mutation \cite{Banyai16}:

\begin{itemize}
\item Mutation by substitution: replacement of one (or more) nucleotides by another (or more)
\item Mutation by insertion: addition of one (or more) nucleotides
\item Mutation by deletion: loss of one (or more) nucleotides
\item Mutation by inversion: permutation of two neighbouring deoxyribonucleotides 
\end{itemize}

These mutations have consequences at the protein level, such as the replacement of one or more amino acids. In addition, the mutation of an amino acid into another with different properties can affect not only the structure of the protein but also its activity. Modification of the protein sequence may also, in some cases, have no effect on its function \cite{Hanna05}.

%========================================================================================
\subsection{Substitution matrices}
%========================================================================================
One of the aims of bioinformatics is to deduce regions of homology within a common ancestral sequence \cite{Xiong2006}. It is important to specify that a pair of amino acids is homologous if the position in the two sequences is identical to that in the common ancestral sequence. Furthermore, if two sequences diverge from a common ancestor, mismatches can be caused by mutations and gaps can be interpreted as indels (insertion or deletion mutations). To assess these transformations, a $20 \times 20$ matrix is constructed in which the coefficient $(i,j)$ is equal to the probability that the $i^{th}$ amino acid will be mutated into the $j^{th}$ amino acid. This type of matrix is known as substitution matrices. The most commonly used are PAM (Point Accepted Mutation) and BLOSUM (BLOck SUbstitution Matrix) matrices.

\begin{figure}[h!]
\centering
\begin{subfigure}{0.49\textwidth}
\centering
\includegraphics[width=0.99\textwidth]{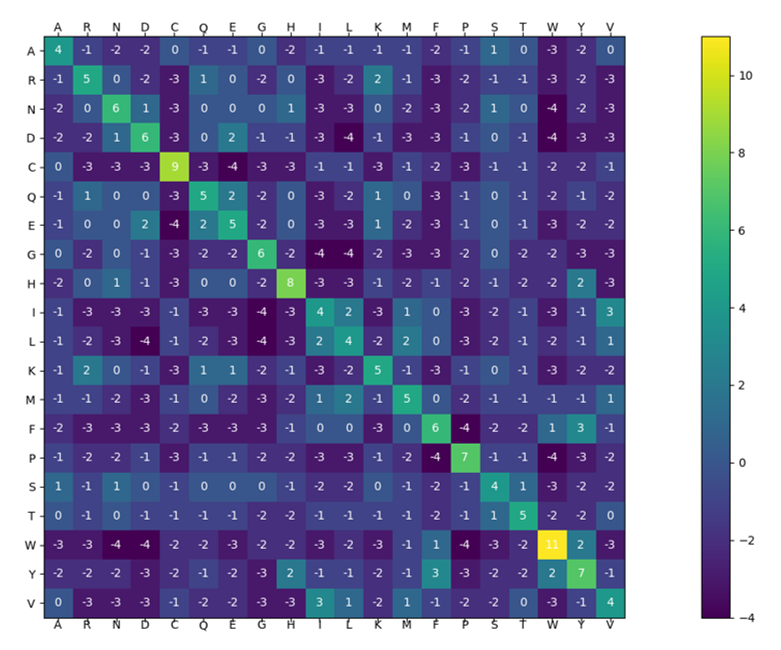}
\caption{BLOSUM62}
\label{Blossum62}
\end{subfigure}
\begin{subfigure}{0.49\textwidth}
\centering
\includegraphics[width=0.99\textwidth]{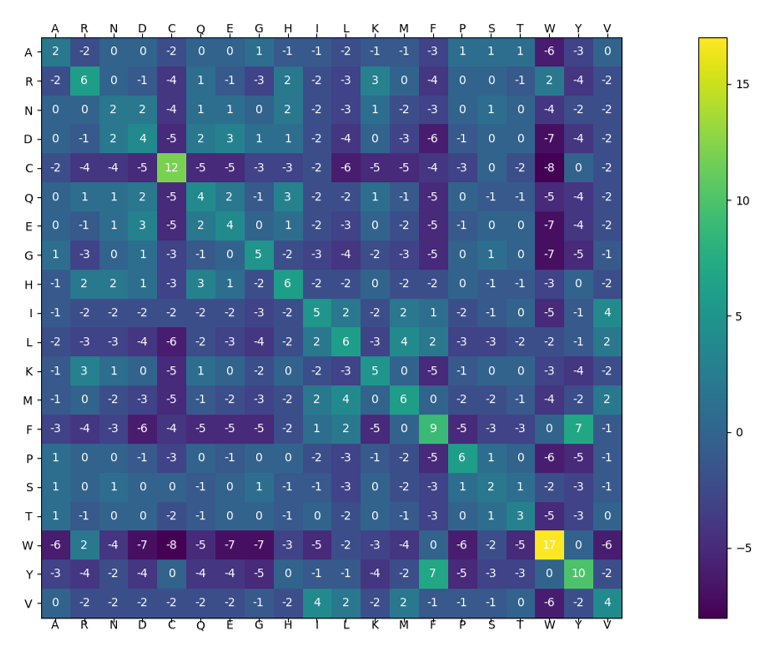}
\caption{PAM250}
\label{Pam250}
\end{subfigure}
\caption{Example of substitution matrices}
\label{Inde}
\end{figure}

The PAM (Point Accepted Mutation) matrix is constructed by observing the differences between proteins with a high degree of similarity. Point mutations can give rise to substitutions that are considered acceptable as a result of natural selection \citep{Dayhoff1978}. A PAM unit is defined as the 1\% of amino acids present in the amino-acids sequences whose locations have been changed. The substitution rate, which would be expected if 1\% of the amino acids present in the protein sequences had been modified, is estimated by a matrix based on a PAM unit and is denoted PAM1 \citep{Dayhoff1978,Altschul2008}. This matrix is used as a basis for calculating other substitution matrices, assuming that repeat mutations follow the same process as those in the PAM1 matrix and that successive substitutions can occur at the same position. Following this logic, matrices up to PAM250 have been built (Figure \ref{Pam250}). PAM30 and PAM70 are the most commonly used \cite{Altschul2008}.

BLOSUM (BLOck SUbstitution Matrix) are substitution matrices developed from ungapped alignment (block), derived from evolutionary divergent sequences of protein \cite{Henikoff1992}. The probabilities, which are used to define the matrix, are calculated by considering the amino-acids blocks that are conserved in the sequences and found in the alignments of a multitude of proteins. These so-called conserved sequences are thought to be crucial in defining the protein's function. They will therefore have lower substitution rates (coefficients in the matrix) than less conserved regions. To reduce the bias of closely related sequences in relation to substitution rate, the segments of a block that are similar in relation to a certain threshold are grouped together in order to reduce the weight of each of these segments \cite{Altschul2008,Trivedi2020,Henikoff1992}. This threshold was set at 62\% for the construction of the BLOSUM62 matrix (Figure \ref{Blossum62}) \cite{Henikoff1992}. A BLOSUM matrix with a high threshold is more likely to be used for two closely related sequences, whereas a lower threshold would be more suitable for more divergent sequences. It turns out that the BLOSUM62 matrix does an excellent job of detecting similarities between distant sequences, and is the default matrix used in most recent alignment applications such as BLAST.

The differences between the PAM and BLOSUM matrices are as follows \citep{Mount2008}:

\begin{itemize}
\item PAM matrices are based on an explicit evolutionary model, while BLOSUM matrices are based on an implicit evolutionary model.
\item PAM matrices include both highly conserved regions and regions with numerous mutations. BLOSUM matrices are based solely on highly conserved regions in a series of alignments where shifts are prohibited.
\item Mutations are not counted in the same way in PAM matrices, unlike the BLOSUM matrix in which mutations are weighted.
\item High coefficients in the PAM matrices indicate a low chance of amino-acids mutation, while high coefficients in the BLOSUM matrices indicate a high chance of mutation and thus an evolutionary trait.
\end{itemize}

BLOSUM62 offers a good compromise when the evolutionary distances between sequences are not known, and this is the matrix used by default in BLAST. This substitution matrix has therefore been chosen for comparison with our reference, i.e., BLAST.

%========================================================================================
\subsection{Dotplot matrix}
%========================================================================================
To visualize the similarity between two protein or nucleic acid sequences, we can use a similarity matrix, known as a dot plot \cite{Gibbs1970}. The protein sequences being compared lie along the vertical and horizontal axes represented by the columns and rows of the matrix respectively. For a simple visual representation of the similarity between two sequences, cells in the matrix can be shaded black if the amino acids are identical, so that sequence segments with similar amino acids appear as diagonal lines running through the matrix.

\begin{figure}[h!]
\centering
\begin{subfigure}{0.3\textwidth}
\centering
\includegraphics[width=0.8\textwidth]{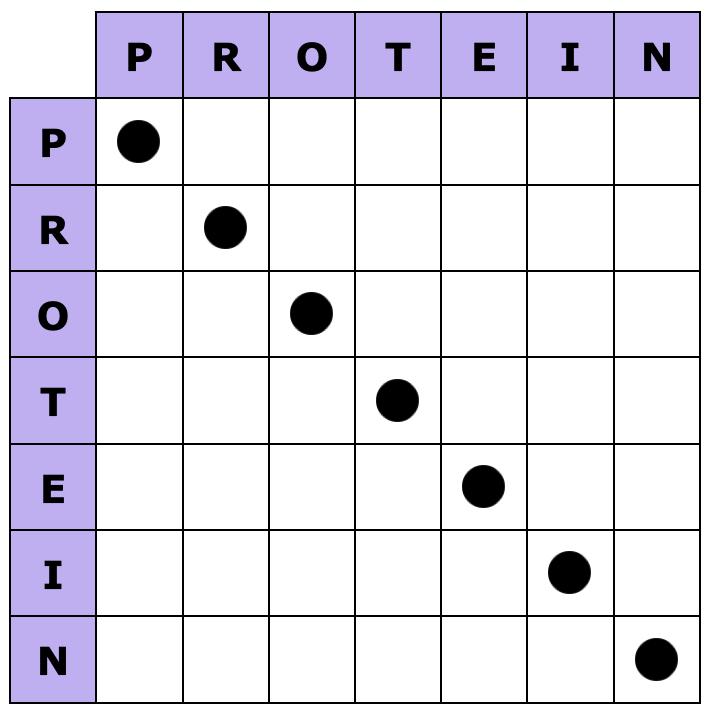}
\caption{Identity}
\label{Identity}
\end{subfigure}
\begin{subfigure}{0.3\textwidth}
\centering
\includegraphics[width=0.8\textwidth]{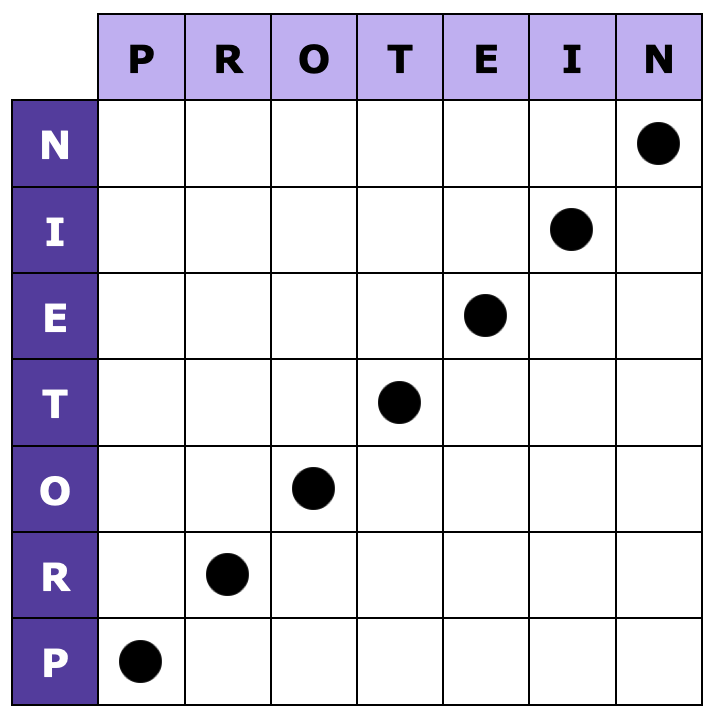}
\caption{Inversion}
\label{Inversion}
\end{subfigure}
\begin{subfigure}{0.33\textwidth}
\centering
\includegraphics[width=0.99\textwidth]{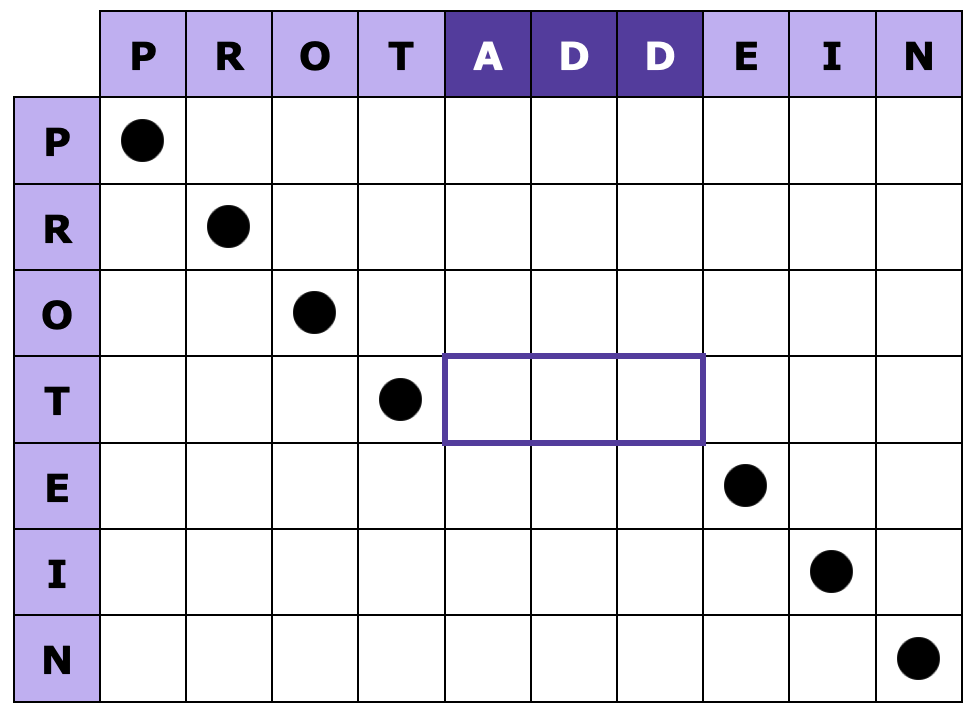}
\caption{Insertion}
\label{Insertion}
\end{subfigure}
\begin{subfigure}{0.25\textwidth}
\centering
\includegraphics[width=0.6\textwidth]{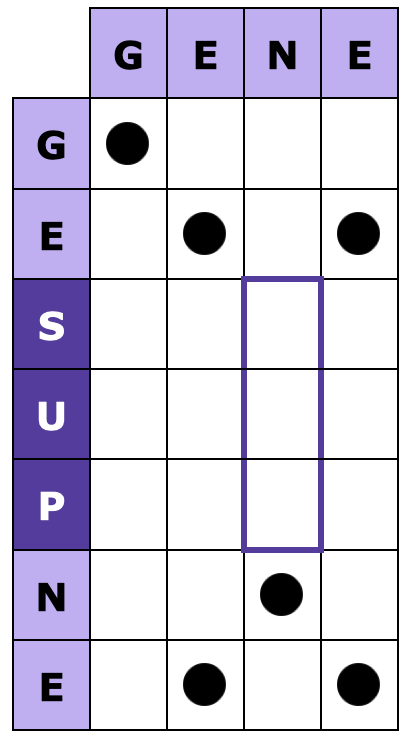}
\caption{Deletion}
\label{Deletion}
\end{subfigure}
\begin{subfigure}{0.33\textwidth}
\centering
\includegraphics[width=0.85\textwidth]{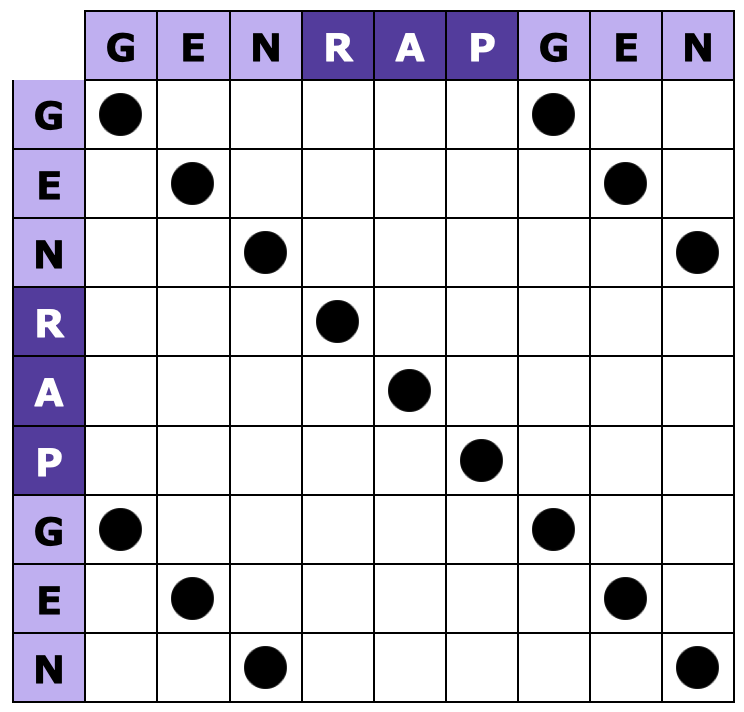}
\caption{Repetition}
\label{Repetition}
\end{subfigure}
\begin{subfigure}{0.33\textwidth}
\centering
\includegraphics[width=0.85\textwidth]{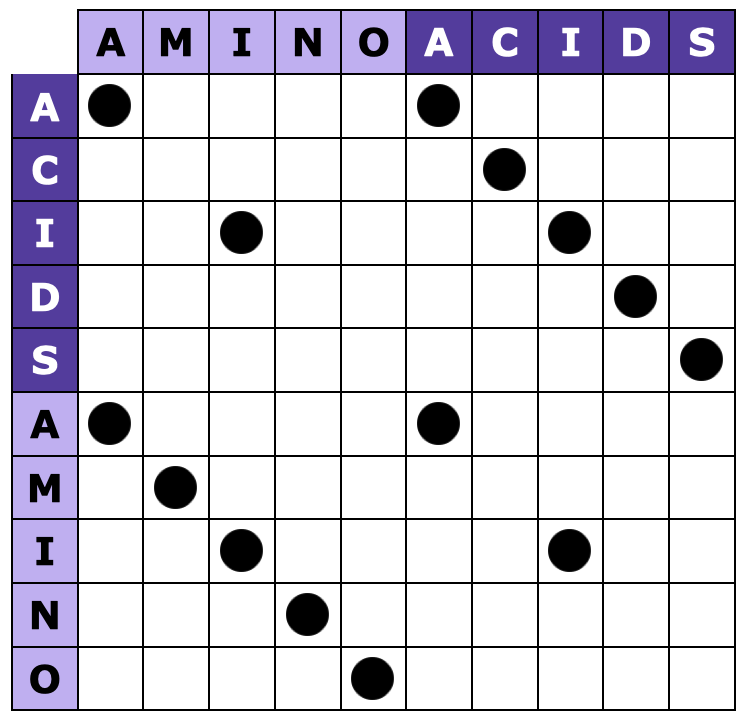}
\caption{Transposition}
\label{Transposition}
\end{subfigure}
\caption{Phenomenon present in the protein amino-acids sequences}
\label{Inde}
\end{figure}

The length of these segments gives an overall idea of the similarity between two proteins sequences. Identical proteins will be represented by the main diagonal of the matrix shaded in black (Figure \ref{Identity}). Insertions, deletions and transpositions between sequences disrupt this representation. Regions of local similarity or repetitive sequences give rise to another diagonal correspondences in addition to the main diagonal.

In graphics, one sequence represents the X-axis and another the Y-axis. When an amino acid is in the same position in both protein sequences, a dot is drawn. Once all the dots have been drawn, we can see that they form lines. The proximity of the sequences in terms of similarity will determine how long the line is.
 
This relationship is influenced by certain features of the sequences, such as frame shifts, direct repeats or inverted repeats. Frame shifts include insertions, deletions and mutations. The presence of one or more of these features will result in several lines being drawn in various possible configurations, depending on the features present in the protein sequences.
 
  It is important to note that this type of matrix  presents inversions (Figure \ref{Inversion}), insertions (Figure \ref{Insertion}), deletions (Figure \ref{Deletion}), repetitions (Figure \ref{Repetition}) or transpositions (Figure \ref{Transposition}) in an easily understandable way \cite{Cabanettes2018}.

For computation, we transform dotplot graph in dot plot matrices. These matrices are constructed with a coefficient of 1 when there is a dot in the graph, 0 otherwise. Using dotplot matrices, we can easily visualise amino-acids sequences, but we cannot take into account the evolutionary character of proteins. To do this, the similarity matrices (dotplot) need to be combined in a process with the substitution matrices (PAM and BLOSUM).

%========================================================================================
\section{Protein quantum generator}
%========================================================================================
The aim of this part is to propose a method to build a sequence in a random way using a quantum algorithm. To determine the probability of an amino acid at a given position, a frequency of occurrence matrix is defined from the Uniprot database.

%========================================================================================
\subsection{Occurrence matrix of amino acids}
%========================================================================================
It is common to model the evolution over time of biological systems using random models, such as an organism or an ecosystem. Here, the method is based on random dynamics and more precisely on Markov chains. To define a Markov chain, it is necessary to have two objects:

\begin{itemize}
\item A known space of states, noted $SP$
\item A transition matrix of size $n \times m$ that will group together probabilities such as,

\begin{equation}
P =\begin{pNiceMatrix}[first-row,first-col]
  & 0 & 1 & \cdots & j & \cdots & m-1  \\
0 & P_{0,0} & P_{0,1} & \cdots & P_{0,j} & \cdots & P_{0,m-1} \\
1 & P_{1,0} & P_{1,1} & \cdots & P_{1,j} & \cdots & P_{1,m-1} \\
\vdots & \vdots & \vdots & \ddots & \vdots & \ddots & \vdots \\
i & P_{i,0} & P_{i,1} & \cdots & P_{i,j} & \cdots & P_{i,m-1} \\
\vdots & \vdots & \vdots & \ddots & \vdots & \ddots & \vdots  \\
n-1 & P_{n-1,0} & P_{n-1,1} & \cdots & P_{n-1,j} & \cdots & P_{n-1,m-1}
\end{pNiceMatrix}
\end{equation}
\end{itemize}

\noindent with,

\begin{align}
& \sum_{j} P_{i,j} = 1 , & \forall \mbox { } 0 \leq i \leq n-1. 
\label{sumprob}
\end{align}
The matrix $P$ is a stochastic rectangular matrix using the equation (\ref{sumprob}).

 In our case, the space of states is the set of different amino acids. As there are 20 of them, we need $ \lfloor\log_2(20) \rfloor + 1 = 5 $ qubits, where $\lfloor . \rfloor$ is the floor function. The states of our quantum system will then go from $\vert 00000 \rangle$ to $\vert 11111 \rangle$. The correlation between the amino acids and the quantum states are shown in Table  \ref{QuantumAminoacid}.

\begin{table}[h!]
\centering
\resizebox{\textwidth}{!}{%
\begin{tabular}{|l|c|c|c|}
\hline
\rowcolor[HTML]{9698ED}  
\textbf{Name} & \textbf{Symbol} & \textbf{Quantum state} \\ \hline 
\rowcolor[HTML]{EBE9F5} 
Alanine & A & $\vert 00000 \rangle$  \\ \hline
\rowcolor[HTML]{EBE9F5}  
Arginine & R &  $\vert 00001 \rangle$ \\ \hline 
\rowcolor[HTML]{EBE9F5} 
Asparagine  & N &  $\vert 00010 \rangle$ \\ \hline 
\rowcolor[HTML]{EBE9F5} 
Aspartic acid  & D &  $\vert 00011 \rangle$ \\ \hline 
\rowcolor[HTML]{EBE9F5} 
Cysteine & C &  $\vert 00100 \rangle$ \\ \hline 
\rowcolor[HTML]{EBE9F5} 
Glutamine & Q &  $\vert 00101 \rangle$ \\ \hline 
\rowcolor[HTML]{EBE9F5} 
Glutamic acid  & E &  $\vert 00110 \rangle$ \\ \hline 
\rowcolor[HTML]{EBE9F5} 
Glycine  & G & $\vert 00111 \rangle$ \\ \hline 
\rowcolor[HTML]{EBE9F5} 
Histidine  & H &  $\vert 01000 \rangle$ \\ \hline 
\rowcolor[HTML]{EBE9F5} 
Isoleucine  & I  & $\vert 01001 \rangle$ \\ \hline 
\end{tabular}
\begin{tabular}{|l|c|c|c|}
\hline
\rowcolor[HTML]{9698ED}  
\textbf{Name} & \textbf{Symbol} & \textbf{Quantum state} \\ \hline 
\rowcolor[HTML]{EBE9F5} 
Leucine  & L &  $\vert 01010 \rangle$ \\ \hline 
\rowcolor[HTML]{EBE9F5} 
Lysine  & K &  $\vert 01011 \rangle$ \\ \hline 
\rowcolor[HTML]{EBE9F5} 
Methionine  & M  & $\vert 01100 \rangle$ \\ \hline 
\rowcolor[HTML]{EBE9F5} 
Phenylalanine  & F & $\vert 01101 \rangle$  \\ \hline 
\rowcolor[HTML]{EBE9F5} 
Proline  & P & $\vert 01110 \rangle$  \\ \hline 
\rowcolor[HTML]{EBE9F5} 
Serine  & S & $\vert 01111 \rangle$  \\ \hline 
\rowcolor[HTML]{EBE9F5} 
Threonine  & T & $\vert 10000 \rangle$  \\ \hline 
\rowcolor[HTML]{EBE9F5} 
Tryptophan  & W & $\vert 10001 \rangle$  \\ \hline 
\rowcolor[HTML]{EBE9F5} 
Tyrosine  & Y & $\vert 10010 \rangle$  \\ \hline 
\rowcolor[HTML]{EBE9F5} 
Valine  & V & $\vert 10011 \rangle$ \\ \hline
\end{tabular}
}
\caption{Correspondence list between amino acids and quantum states}
\label{QuantumAminoacid}
\end{table}

In our cases, the two necessary objects are:

\begin{itemize}
\item The set of amino acids that will be our known states space of finite dimension, 

\begin{equation}
SP := \lbrace \mbox{A,R,N,D,C,Q,E,G,H,I,L,K,M,F,P,S,T,W,Y,V}  \rbrace.
\end{equation}

\item A transition matrix that will group together the probability of occurrence relative to the different positions and amino acids such as,

\begin{align}
P &= Prob(Position = i, Amino Acid = j)_{0 \leq i \leq x-1, 0 \leq j \leq 19} \nonumber \\ &=\begin{pNiceMatrix}[first-row,first-col]
  & \mbox{A} & \mbox{R} & \cdots & \cdots & \cdots & \mbox{V}  \\
0 & P_{0,0} & P_{0,1} & \cdots & P_{0,j} & \cdots & P_{0,19} \\
1 & P_{1,0} & P_{1,1} & \cdots & P_{1,j} & \cdots & P_{1,19} \\
\vdots & \vdots & \vdots & \ddots & \vdots & \ddots & \vdots \\
i & P_{i,0} & P_{i,1} & \cdots & P_{i,j} & \cdots & P_{i,19} \\
\vdots & \vdots & \vdots & \ddots & \vdots & \ddots & \vdots  \\
x-1 & P_{x-1,0} & P_{x-1,1} & \cdots & P_{x-1,j} & \cdots & P_{x-1,19}
\end{pNiceMatrix}
\end{align}
\end{itemize}

\noindent where the sum of the coefficients of each line is equal to 1. This matrix is constructed by listing the probabilities of each amino acid for each position in the sequence of proteins from the database Uniprot. Here, we have a quantum system for each position, i.e., a quantum system for each row of the transition matrix. The states coefficients depend on the transition matrix (or matrix of occurrence) and the state of the $i^{th}$ quantum system is:

\begin{align}
& \vert \psi_i \rangle = \sum_{j=0}^{2^5-1} \beta_j \vert j \rangle & \mbox{with } \beta_j = \begin{cases}\sqrt{P_{i,j}} \mbox{ if } j < 20 \\ 0 \mbox{ else}
\end{cases}.
\end{align}

%========================================================================================
\subsection{Quantum circuit}
\begin{figure}[h!]
\centering
\includegraphics[width=0.9\textwidth]{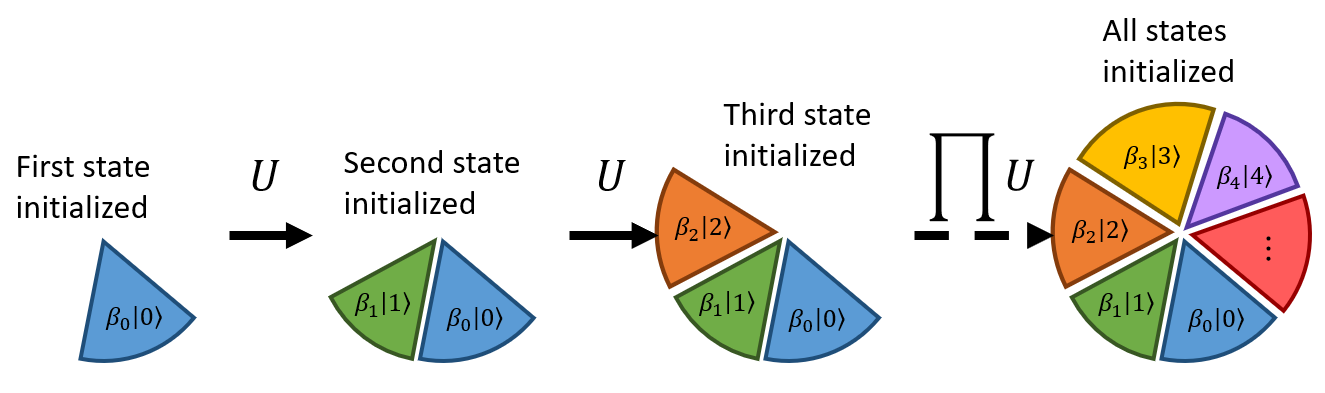}
\caption{Cake algorithm}
\label{cake}
\end{figure}

%========================================================================================
To initialize the quantum state in the circuit, we'll use the cake algorithm, i.e., the coefficients of the basic states of the quantum system will be initialized as we go along, in order to go through all the states. To do this, we need two registers, the first corresponding to the state to be initialized (in other words the amino acid), with a size of $n_{aa} = 5$ qubits, and the second, which is only used to orient the measurement and obtain the correct coefficient (and above all the correct sign) on the last base state, with a size of $n_\lambda = 1$ qubit. The quantum system evolves as follows,

\begin{equation}
\lvert 0 \rangle_{n_{aa} +n_\lambda} \Rightarrow \sum_{j = 0}^{2^{n_{aa}-1}}   \beta_j \lvert j  \rangle_{n_{aa}}  \vert 0 \rangle_{n_\lambda} + r_{aux} \lvert 2^{n_{aa}}-1  \rangle_{n_{aa}}  \vert 1 \rangle_{n_\lambda} = \begin{pmatrix}   \beta_0 \\   \beta_1 \\   \beta_2 \\ \vdots \\   \beta_{2^{n_{aa}}-1} \\ 0 \\ \vdots \\ 0 \\ r_{aux} \end{pmatrix}, 
\end{equation}

\noindent with $\beta_j$ the coefficients to be initialized and $r_{aux}$ the coefficient remains. Only those states where the register $n_\lambda$ is set to $\vert 0 \rangle_{n_\lambda}$correspond to the states to be initialized. For each initialisation, assuming that a gate, called $U$, allows initialisation, the complete process corresponding to the calculations is as follows,

\begin{align*}
\lvert 0 \rangle_{n_{aa}+n_\lambda} &  \substack{ U \\ \Rightarrow} \text{ } \beta_0 \lvert 0 \rangle_{n_{aa}+n_\lambda} +  r_0 \lvert \overline{\psi_0} \rangle_{n_{aa}+n_\lambda} \\
&  \substack{ U \\ \Rightarrow}  \text{ } \beta_0 \lvert 0 \rangle_{n_{aa}+n_\lambda} +  \beta_1 \lvert 1 \rangle_{n_{aa}+n_\lambda} + r_1 \lvert \overline{\psi_1} \rangle_{n_{aa}+n_\lambda} \\
&  \substack{ U \\ \Rightarrow} \text{ }  \beta_0 \lvert 0 \rangle_{n_{aa}+n_\lambda} +   \cdots +  \beta_2 \lvert 2 \rangle_{n_{aa}+n_\lambda} + r_2 \lvert \overline{\psi_2}
\rangle_{n_{aa}+n_\lambda} \\
\vdots \\
&  \substack{ U \\ \Rightarrow} \text{ }  \beta_0 \lvert 0 \rangle_{n_{aa}+n_\lambda} +  \cdots +  \beta_{2^{n_{aa}-1}} \lvert 2^{n_{aa}-1}  
\rangle_{n_{aa}+n_\lambda} +r_{aux} \vert aux \rangle_{n_{aa}+n_\lambda} \\
&  \substack{ U \\ \Rightarrow } \text{ }  \beta_0 \lvert 0 \rangle_{n_{aa}}\lvert 0 \rangle_{n_\lambda} +  \cdots +  \beta_{2^{n_{aa}-1}} \lvert 2^{n_{aa}-1}  
\rangle_{n_{aa}}\lvert 0 \rangle_{n_\lambda} +r_{aux} \vert 2^{n_{aa}-1} \rangle_{n_{aa}}\lvert 1\rangle_{n_\lambda} \\
& \Leftrightarrow \sum_{j = 0}^{2^{n_{aa}-1}}  \beta_j \lvert j \rangle_{n_{aa}}  \vert 0 \rangle_{n_\lambda} + r_{aux} \lvert 2^{n_{aa}}-1  \rangle_{n_{aa}}  \vert 1 \rangle_{n_\lambda}   
\end{align*}

\noindent where $\vert \overline{\psi_j} \rangle_{n_{aa}+n_\lambda}$ are the states yet to be initialized. The computation can be schematically illustrated as shown in Figure \ref{cake}.

All that remains is to determine the $U$ gate in question. This gate must influence only two basic quantum states. In addition, the measurement probability is modified, i.e., there is a rotation around the Y-axis in the Blotch sphere. The $U$ gate will be the multi-controlled $Ry$ gate with an angle $\theta$ as argument. The various angles $\theta_j$ are defined as follows,

\begin{equation}
\theta_j  = 2 \arccos \left( \frac{  \beta_j}{\sqrt{1- \sum_{k = 0}^{j-1} \beta_k^2 }} \right) = \begin{cases} 2 \arccos \left( \frac{  \sqrt{P_{i,j}}}{\sqrt{1- \sum_{k = 0}^{j-1} P_{i,k} }} \right) \mbox{ if } j < 20 \\ 0 \mbox{ else }
\end{cases}.
\end{equation}

\begin{figure}[h!]
\centering
\includegraphics[width=0.9\textwidth]{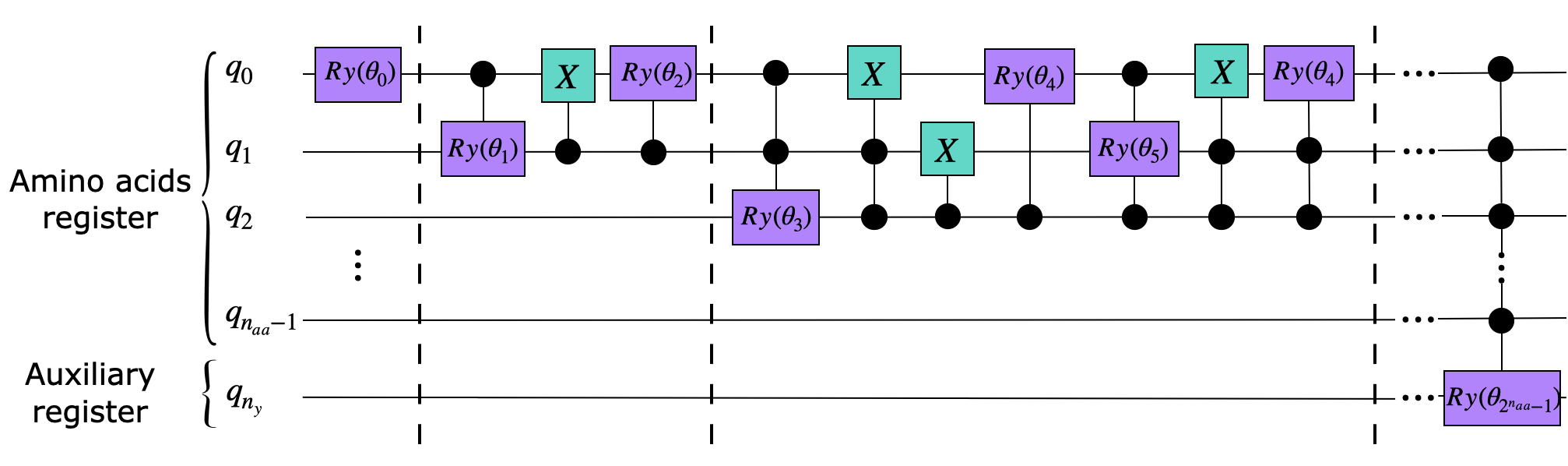}
\caption{Representative circuit of the cake algorithm for eight amino acids}
\label{Circ_cake}
\end{figure}

In order to target the current quantum state, multi-controlled X-gates are added to the circuit. The quantum circuit of the initialisation of eight amino acids follows an iterative construction procedure and represented in Figure \ref{Circ_cake}. 

\begin{figure}[h!]
\centering
\includegraphics[width=0.9\textwidth]{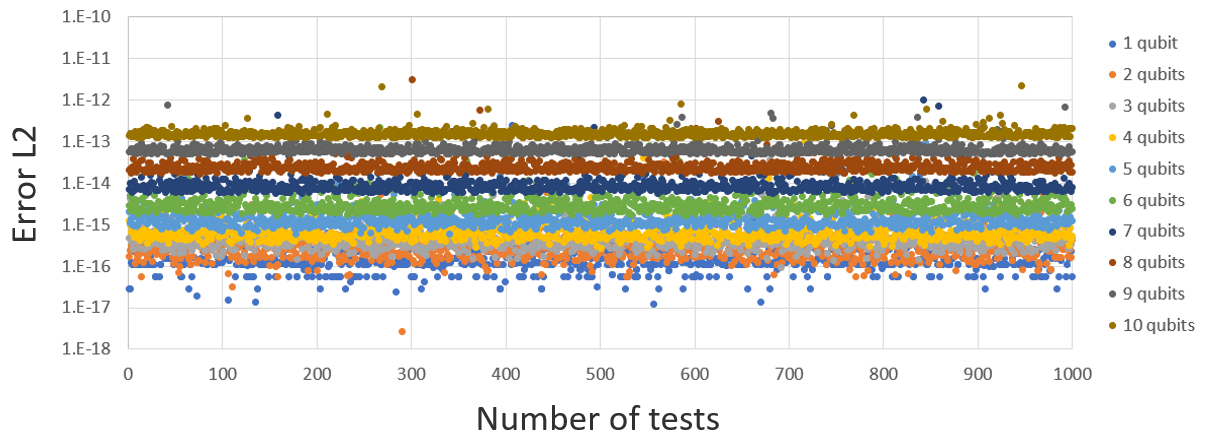}
\caption{Error $L^2$ of initialisation of a quantum state}
\label{Error_init}
\end{figure}

To test the initialisation, we compared the output state of the quantum circuit with the initial vector for 1000 systems ranging from one to ten qubits and calculated the $L^2$ error. Figure \ref{Error_init} shows that the error is between $10^{-10}$ and $10^{-14}$, so the process does initialise the quantum circuit to the given state.

%========================================================================================
\section{First method for protein similitude measurement: Bioinformatics algorithms}
%========================================================================================
In bioinformatics, there are two common algorithms for comparing two sequences of amino acids: Needleman-Wunsch and Smith-Waterman algorithms.

%========================================================================================
\subsection{Needleman-Wunsch algorithm}
%========================================================================================

The Needleman-Wunsch algorithm is one of the first applications of dynamic programming for comparing amino-acids sequences. It is used in bioinformatics to align amino-acids or nucleotides sequences. This algorithm uses a \textquotedblleft{} divide and conquer \textquotedblright{} method. In other words, the large problem (in this case, the complete amino-acids sequence) will be subdivided into a series of smaller problems (smaller sequences). The particular solutions to the smaller problems will be used to find a global solution to the larger problem \cite{Needleman1970,Gagniuc2021}. This algorithm is also sometimes referred to as a global alignment algorithm. The algorithm can be divided into two steps: 

\begin{itemize}
\item Creation of a scoring based on a chosen scoring system
\item Traceback to find the optimal alignment between two protein sequences
\end{itemize}

The algorithm calculates a score, according to the chosen system, for each possible alignment found. The aim is then to find all the possible alignments with the highest score.

%========================================================================================
\subsubsection{Needleman-Wunsch scoring matrix}
%========================================================================================
The first step is to construct a matrix, denoted $M$, where the amino acids of the first protein (sequence $A$) is represented by the columns and the amino acids of the second protein (sequence $B$) is represented by the rows. A row and a column are added at the top and left of the matrix respectively for initialisation, in order to align the sequences correctly and to take account of the gap penalty at initialisation. Before filling in the matrix, a scoring system must be chosen. Different configurations may be encountered. The amino acids may match, mismatch or correspond to a gap (a deletion or insertion). For each of these configurations, a score is assigned and the sum of the scores for all the amino-acids pairs is the total score for the entire candidate alignment. Amino acid matches and mismatches values are defined by the substitution matrix, in our case the BLOSUM62 matrix. For deletions and insertions, a gap penalty is defined. For this penalty, we take a linear penalty, i.e., the same penalty is considered for the opening and extension of a gap. The gap penalty is denoted $GP_k$ with $k$ the length of the gap and is defined by:

\begin{equation}
GP_k = k GP_1,
\end{equation}

\noindent where $GP_1$ is the cost of a single gap. In the initialisation the first row and column of the scoring matrix are defined as,

\begin{equation}
\begin{cases}
M_{0,j} = - j GP_1 \\
M_{i,0} = - i GP_1
\end{cases}.
\end{equation}

For the other matrix coefficients, they are defined by recursion based on the principle of optimality as for a linear gap penalty,

\begin{equation}
M_{i,j} = \max \begin{cases}
M_{i-1,j-1} + \mbox{BLOSUM62} (A_i,B_j) \\
M_{i-1,j} - GP_1 \\
M_{i,j-1} - GP_1
\end{cases}.
\end{equation}

%========================================================================================
\subsubsection{Needleman-Wunsch traceback}
%========================================================================================
For the traceback, the starting point is the last coefficient of the scoring matrix. The coefficients of this matrix are then traversed until the first row or column is reached. During the traceback, the largest coefficient in each case is considered when moving up the matrix. If there is a choice of several coefficients, these represent a branching of the alignments. If two or more branches all belong to alignments with the same start and end cells, these are equally viable alignments. In this case, the different alignments are noted as distinct candidates. To represent the traceback graphically, arrows will be used to construct a sequence, two types of arrows can result:

\begin{itemize}
\item A diagonal arrow represents a match or mismatch.
\item A horizontal or vertical arrow represents an insertion in one of the two sequences. Horizontal arrows represent an insertion in sequence $A$ and vertical arrows represent an insertion in sequence $B$.
\end{itemize}

\begin{figure}[h!]
\centering
\includegraphics[width=0.9\textwidth]{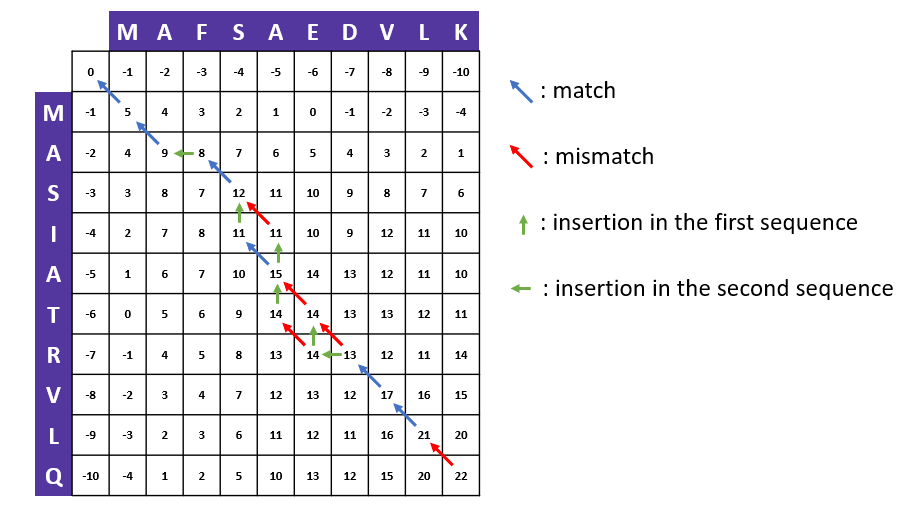}
\caption{Scoring matrix and traceback for the Needleman-Wunsch algorithm}
\label{NW_traceback}
\end{figure}

For example, with a first sequence MAFSAEDVLK and a second sequence MASIATRVLQ, the scoring matrix (with $GP_1 = 1$) and traceback, for the Needleman-Wunsch algorithm, are represented Figure \ref{NW_traceback}.

%========================================================================================
\subsection{Smith-Waterman algorithm}
%========================================================================================
Like the Needleman-Wunsch algorithm, the Smith-Waterman algorithm perform a sequence alignment, but contrary to the Needleman-Wunsch algorithm where the alignment is global. This time, the alignment is local; that is, for determining similar regions between two protein sequences. The Smith-Waterman algorithm takes a more local approach. Instead of examining the entire sequence, it compares segments of all possible lengths to obtain a more accurate similarity rate. As such, it has the property of guaranteeing that it finds the optimal local alignment with regard to the scoring system chosen (which includes the substitution matrix and the gap scoring system). This makes the local alignments more visible (and therefore gives them a positive score) than the Needleman-Wunsch algorithm \cite{Gagniuc2021,Smith1981}.

%========================================================================================
\subsubsection{Smith-Waterman scoring matrix}
%========================================================================================
The construction of the matrix is very similar for both algorithms, only the initialisation and the traceback differ. In the initialisation, the first row and column of the scoring matrix no longer depend on the gap penalty and are defined as,

\begin{equation}
\begin{cases}
M_{0,j} = 0 \\
M_{i,0} = 0
\end{cases}.
\end{equation}

For the other coefficients of the matrix, they are defined by recursion based on Bellman's optimality principle \citep{Dreyfus02}, for a linear gap penalty, and now all the coefficients are defined as positive,

\begin{equation}
M_{i,j} = \max \begin{cases} 
M_{i-1,j-1} + \mbox{BLOSUM62} (A_i,B_j) \\
M_{i-1,j} - GP_1 \\
M_{i,j-1} - GP_1 \\
0
\end{cases}.
\end{equation}

%========================================================================================
\subsubsection{Smith-Waterman traceback}
%========================================================================================
For the traceback of the Smith-Waterman algorithm, its starting point is the highest coefficient of the scoring matrix and continues until a cell with a value of zero is reached. As with the Needleman-Wunsch algorithm, we have the same configurations, i.e., the types of arrows can result. To recall the two types of arrows, we have:

\begin{itemize}
\item A diagonal arrow represents a match or mismatch.
\item A horizontal or vertical arrow represents an insertion in one of the two sequences. Horizontal arrows represent an insertion in sequence $A$ and vertical arrows represent an insertion in sequence $B$.
\end{itemize}
 
\begin{figure}[h!]
\centering
\includegraphics[width=0.9\textwidth]{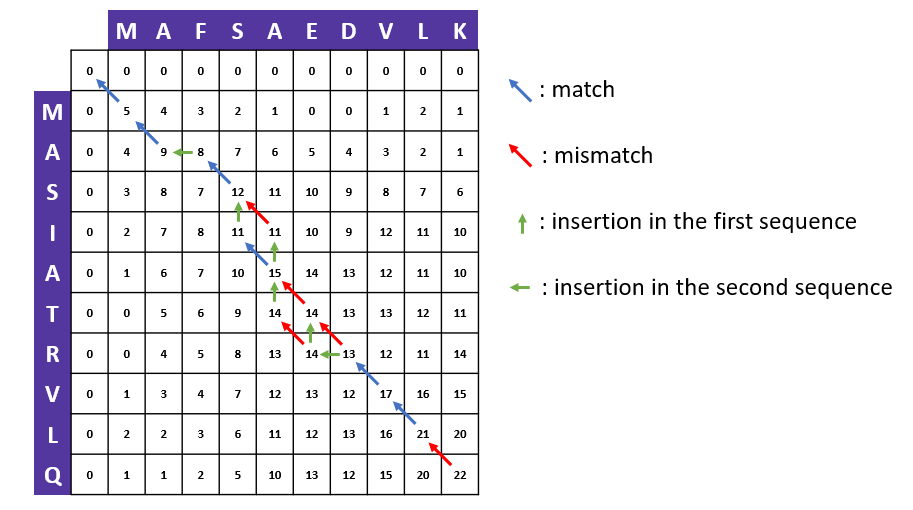}
\caption{Scoring matrix and traceback for the Smith-Waterman algorithm}
\label{SW_traceback}
\end{figure}

The aim of the Smith-Waterman algorithm is the same as that of Needleman-Wunsch, i.e., to find candidates for sequence alignment. Its traceback allows us to obtain the alignment with the highest similarity score in relation to the scoring system chosen. To obtain the second best local alignment, simply repeat the search, but this time starting from the second highest score outside the trace of the first alignment. For example, for the Smith-Waterman algorithm, with the same sequences, the scoring matrix and traceback are shown in Figure \ref{SW_traceback}.

%========================================================================================
\subsection{Protein sequence alignment}
%========================================================================================
In bioinformatics, sequence alignment represents a fundamental procedure by which two or more sequences of DNA, RNA and/or proteins are compared to identify regions of similarity. To match the equivalent positions of amino acids, gaps are inserted in the sequences. The sequence alignment indicates the positions of insertions and deletions and allow us to identify important functional motifs \cite{Higgs2004}. As we will be comparing two sequences each time, the procedure is called the pairwise alignment. To find the sequence alignment, we traceback the bioinformatics algorithms.

For example, with the previous same sequence, we can have the two following results, shown in Table  \ref{Pairwise}, same for the two algorithms (Needleman-Wunsch and Smith-Waterman). We've kept the same color code as before ({\color[HTML]{4472C4} \textbf{blue}} for a match, {\color[HTML]{FE0000} \textbf{red}} for a mismatch and {\color[HTML]{70AD47} \textbf{green}} for a gap).

\begin{table}[h!]
\centering
\begin{subtable}{0.45\textwidth}
\centering
\resizebox{\textwidth}{!}{
\begin{tabular}{ccccccccccc}
{\color[HTML]{4472C4} \textbf{M}} & {\color[HTML]{4472C4} \textbf{A}} & {\color[HTML]{70AD47} \textbf{F}} & {\color[HTML]{4472C4} \textbf{S}} & {\color[HTML]{70AD47} \textbf{-}} & {\color[HTML]{4472C4} \textbf{A}} & {\color[HTML]{FE0000} \textbf{E}} & {\color[HTML]{FE0000} \textbf{D}} & {\color[HTML]{4472C4} \textbf{V}} & {\color[HTML]{4472C4} \textbf{L}} & {\color[HTML]{FE0000} \textbf{K}}   \\ 
{\color[HTML]{4472C4} \textbf{M}} & {\color[HTML]{4472C4} \textbf{A}} & {\color[HTML]{70AD47} \textbf{-}} & {\color[HTML]{4472C4} \textbf{S}} & {\color[HTML]{70AD47} \textbf{I}} & {\color[HTML]{4472C4} \textbf{A}} & {\color[HTML]{FE0000} \textbf{T}} & {\color[HTML]{FE0000} \textbf{R}} & {\color[HTML]{4472C4} \textbf{V}} & {\color[HTML]{4472C4} \textbf{L}} & {\color[HTML]{FE0000} \textbf{Q}}
\end{tabular}}
\caption{Close to diagonal result}
\label{Pairwise_diag}
\end{subtable}
\hfill
 \begin{subtable}{0.45\textwidth}
\centering
\resizebox{\textwidth}{!}{
\begin{tabular}{cccccccccccc}
{\color[HTML]{4472C4} \textbf{M}} & {\color[HTML]{4472C4} \textbf{A}} & {\color[HTML]{70AD47} \textbf{F}} & {\color[HTML]{4472C4} \textbf{S}} & {\color[HTML]{70AD47} \textbf{-}} & {\color[HTML]{70AD47} \textbf{-}} & {\color[HTML]{FE0000} \textbf{A}} & {\color[HTML]{FE0000} \textbf{E}} & {\color[HTML]{70AD47} \textbf{D}} & {\color[HTML]{4472C4} \textbf{V}} & {\color[HTML]{4472C4} \textbf{L}} & {\color[HTML]{FE0000} \textbf{K}}   \\ 
{\color[HTML]{4472C4} \textbf{M}} & {\color[HTML]{4472C4} \textbf{A}} & {\color[HTML]{70AD47} \textbf{-}} & {\color[HTML]{4472C4} \textbf{S}} & {\color[HTML]{70AD47} \textbf{I}} & {\color[HTML]{70AD47} \textbf{A}} & {\color[HTML]{FE0000} \textbf{T}} & {\color[HTML]{FE0000} \textbf{R}} & {\color[HTML]{70AD47} \textbf{-}} & {\color[HTML]{4472C4} \textbf{V}} & {\color[HTML]{4472C4} \textbf{L}} & {\color[HTML]{FE0000} \textbf{Q}}
\end{tabular}}
\caption{Off-diagonal results}
\label{Pairwise_off}
\end{subtable}
\caption{Pairwise sequence alignment example}
\label{Pairwise}
\end{table}

First, we calculate the score between the test protein (with $N_t$ amino acids) and itself in order to obtain a reference score, noted as $\Gamma_{ref}$. This reference score is obtained by adding the match value in the substitution matrix (BLOSUM62) and After, we consider the $i^{th}$ protein (with $N_i$ amino acids) of database (Uniprot) and we compute the score between the two proteins, by using the traceback and the sequence alignment considering the $j^{th}$ candidate in the latter, noted as $\Gamma_{i,j}$ and defined as,

\begin{equation}
\begin{split}
\Gamma_{i,j} = &\underbrace{\sum_{k = 1}^{n_{ma}} \mbox{BLOSUM62}(\mbox{Amino Acid}^{match_1}_k,\mbox{Amino Acid}^{match_2}_k)}_{\sum Match} + \\ &\underbrace{\sum_{k = 1}^{n_{mi}} \mbox{BLOSUM62}(\mbox{Amino Acid}^{mismatch_1}_k,\mbox{Amino Acid}^{mismatch_2}_k)}_{\sum Mismatch} + \underbrace{\sum_{k = 1}^{n_{g}} GP_1}_{\sum Gap}
\end{split}
\end{equation}
\noindent where $n_{ma}$ is the match number, $n_{mi}$ is the mismatch number and $n_g$ is the number of gaps.

 The rate of similarity, between the two proteins for the two bioinformatics algorithms for classic version, noted $S_i$, is then defined as follows:

\begin{equation}
S_i = \left( \frac{ \max_{j} \left( \Gamma_{i,j} \right)}{ \Gamma_{ref}} \right) \left( \frac{\min (N_t,N_i)}{\max (N_t,N_i)} \right).
\end{equation} 

The coefficient $\frac{\min (N_t,N_i)}{\max (N_t,N_i)}$ is used to normalise the similarity and the latter is included in $\lbrace 0,1 \rbrace$. This is in order to get an idea of the similarity depending on the size of the proteins to avoid having a high similarity for long amino-acids sequences. For the example, the reference score is,

\begin{align*}
\Gamma_{ref} &= \mbox{BLOSUM62}(\mbox{M},\mbox{M}) + \mbox{BLOSUM62}(\mbox{A},\mbox{A}) + \mbox{BLOSUM62}(\mbox{F},\mbox{F}) + \mbox{BLOSUM62}(\mbox{S},\mbox{S}) \\ &+ \mbox{BLOSUM62}(\mbox{A},\mbox{A}) + \mbox{BLOSUM62}(\mbox{E},\mbox{E}) + \mbox{BLOSUM62}(\mbox{D},\mbox{D}) + \mbox{BLOSUM62}(\mbox{V},\mbox{V}) \\ &+ \mbox{BLOSUM62}(\mbox{L},\mbox{L}) + \mbox{BLOSUM62}(\mbox{K},\mbox{K}) \\
&= 5 + 4 + 6 + 4 + 4+ 5+6+4+4+5\\
&= 47
\end{align*}

For the first sequence alignment (Table \ref{Pairwise_diag}), we have the score,

\begin{align*}
\Gamma_{a} &= \mbox{BLOSUM62}(\mbox{M},\mbox{M}) + \mbox{BLOSUM62}(\mbox{A},\mbox{A}) + \mbox{BLOSUM62}(\mbox{S},\mbox{S}) + \mbox{BLOSUM62}(\mbox{A},\mbox{A}) \\ &+ \mbox{BLOSUM62}(\mbox{E},\mbox{T}) + \mbox{BLOSUM62}(\mbox{D},\mbox{R}) + \mbox{BLOSUM62}(\mbox{V},\mbox{V}) + \mbox{BLOSUM62}(\mbox{L},\mbox{L}) \\ &+ \mbox{BLOSUM62}(\mbox{K},\mbox{Q}) - 2GP_1 \\
&= 5+4+4+4-1-2+4+4+1-2\\
&= 21
\end{align*}

\noindent and for the second sequence alignment (Table \ref{Pairwise_off}), we have the score,

\begin{align*}
\Gamma_{b} &= \mbox{BLOSUM62}(\mbox{M},\mbox{M}) + \mbox{BLOSUM62}(\mbox{A},\mbox{A}) + \mbox{BLOSUM62}(\mbox{S},\mbox{S}) + \mbox{BLOSUM62}(\mbox{A},\mbox{T}) \\ &+ \mbox{BLOSUM62}(\mbox{E},\mbox{R}) + \mbox{BLOSUM62}(\mbox{V},\mbox{V}) + \mbox{BLOSUM62}(\mbox{L},\mbox{L}) + \mbox{BLOSUM62}(\mbox{K},\mbox{Q}) - 4GP_1 \\
&= 5+4+4+0+0+4+4+1-4\\
&= 18
\end{align*}

We can now compute the similarity between the two proteins,

\begin{align*}
S &= \left( \frac{ \max(\Gamma_{a}, \Gamma_b)}{ \Gamma_{ref}} \right) \left( \frac{\min (10,10)}{\max (10,10)} \right)\\
&= \frac{21}{47} = 0.447
\end{align*} 

Between these two sequences, we have a similarity of $44.7 \%$.

%========================================================================================
\subsection{Comparison between the two algorithms}
%========================================================================================
The Needleman-Wunsch algorithm attempts to align every residue within the entire sequences. The final purpose is to ensure the sequentially homologous aspect of the aligned sequences. This type of alignment is most useful if the sequences in the query set are similar and of roughly equal size. On the other hand, the Smith-Waterman algorithm is more useful for dissimilar and divergent sequences since it identifies regions of similarity regardless their position within long sequences. Both algorithms use the same construction: a substitution matrix, a deviation penalty function, a scoring matrix and a tracing process. Three main differences are \citep{Parvez20}:

\begin{itemize}
\item Initialisation: For Needleman-Wunsch algorithm, the first row and column are subject to gap penalty, whereas for Smith-Waterman algorithm, the first row and column are set to 0.
\item Score: For Needleman-Wunsch algorithm, the score can be negative, whereas for Smith-Waterman algorithm, negatives scores are set to 0.
\item Traceback : For Needleman-Wunsch algorithm, the beginner is the cell in the lower right of the matrix and the end is when the traceback reaches the first row or column, whereas for Smith-Waterman algorithm, the beginner is the cell with the highest score and the end is when 0 is encountered.
\end{itemize}

One of the most important differences is that no negative score is assigned in the scoring system of the Smith-Waterman algorithm. When an element has a negative score, this means that the amino-acids sequences up to that position have no similarities.  This coefficient in the matrix will then be initialised at zero to eliminate the influence of the previous alignment. In this way, the algorithm can continue to find an alignment in any position thereafter. The initialisation of the score matrix in the Needleman-Wunsch algorithm takes into account the deviation penalty for aligning complete sequences, whereas the initialisation of the score matrix in the Smith-Waterman algorithm with zero values allows any segment of a sequence to be aligned with an arbitrary position in the other sequence. 

\begin{figure}[h!]
\centering
\begin{subfigure}{0.49\textwidth}
\centering
\includegraphics[width=0.99\textwidth]{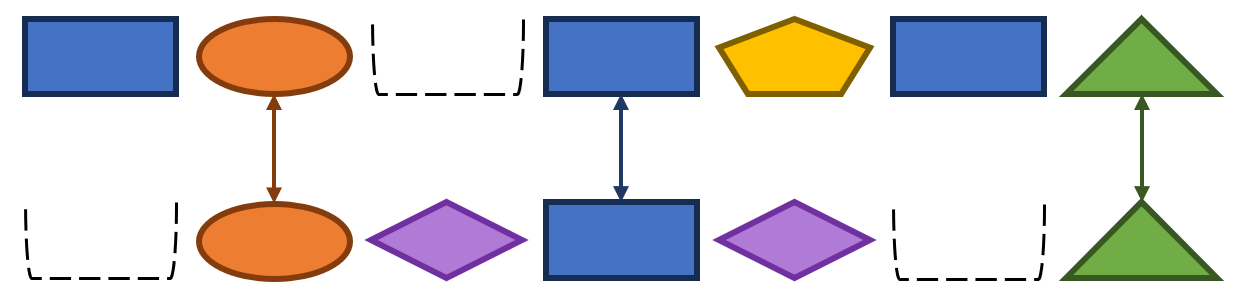}
\caption{Global alignment}
\label{Global_align}
\end{subfigure}
\begin{subfigure}{0.49\textwidth}
\centering
\includegraphics[width=0.99\textwidth]{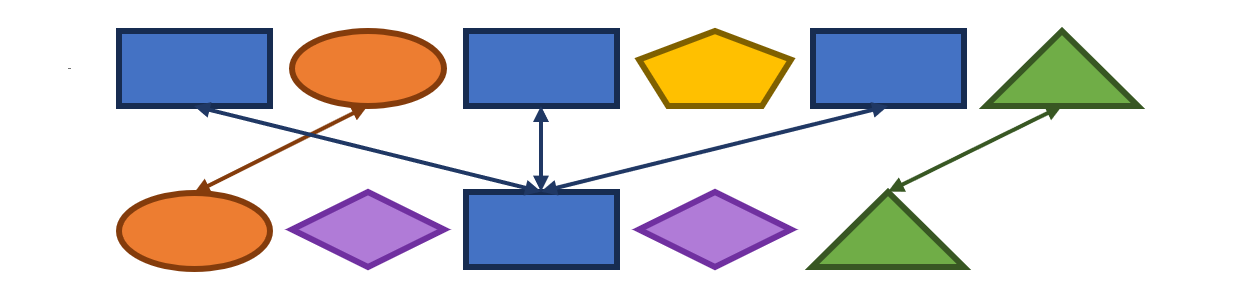}
\caption{Local alignment}
\label{Local_align}
\end{subfigure}
\caption{The two types of alignments in amino-acids sequence}
\label{Align}
\end{figure}

If we consider the dotplot matrix, the global alignment will be represented by the matrix's main diagonal and the local alignments by the diagonals outside the main diagonal. To illustrate the two types of alignment in an amino-acids sequence, global alignment is shown in Figure \ref{Global_align} and local alignment in Figure \ref{Local_align}.

%========================================================================================
\subsection{Quantum version of the two algorithms}
%========================================================================================
Both algorithms (Needleman-Wunsch and Smith-Waterman) have the same structure:

\begin{itemize}
\item Choice of a scoring system 
\item Construction of the scoring matrix
\item Traceback to obtain the score and thus the similarity rate between two proteins
\end{itemize}

The classical and quantum versions of the different algorithms use the same score system and scoring matrix. What will differentiate the two versions is the traceback, which will be replaced by a combinatorial optimisation problem (a model QUBO). In fact, in the quantum versions we are not going to calculate the similarity rate explicitly but the energy rate between the two proteins. First, we calculate the energy ratio between the test protein (with $N_t$ amino acids) and itself in order to obtain a reference ratio, noted as $\Xi_{ref}$. After, we consider the $i^{th}$ protein (with $N_i$ amino acids) of database (Uniprot) and we compute the energy ratio between the two proteins, by solving the optimisation problem for the scoring matrix, noted as $\Xi_{i}$. The rate of similarity between the two proteins, noted $S_i$, is then defined as follows:

\begin{equation}
S_i = \left( \frac{ \Xi_{i}}{ \Xi_{ref}} \right) \left( \frac{\min (N_t,N_i)}{\max (N_t,N_i)} \right).
\end{equation} 

The aim is to carry out this procedure for all the proteins in the database in order to repeat the different similarities and to identify and classify the proteins that are closest to the test protein. 

%========================================================================================
\section{Second method for protein similitude measurement: Conflict graph algorithm}
%========================================================================================
To compare two proteins, we will adapt the algorithm explained in \cite{Jimenez21} for the case of proteins. Here, the aim of the algorithm is to obtain a similarity value between two proteins. Throughout the algorithm, the proteins are represented by graphs where the vertices are the amino acids that make up the proteins and the edges are the peptide bonds. The algorithm can be represented in two steps:

\begin{itemize}
\item Creation of a conflict graph (graph of similarities between proteins)
\item Calculating similarity from the conflict graph (optimisation problem)
\end{itemize}

To create the vertices of the conflict graph, we will browse the amino acids of the two proteins and add a vertex to the conflict graph if the two amino acids share properties in common:

\begin{itemize}
\item Amino-acids symbol
\item Degree of an amino-acid (number of linked amino acids)
\end{itemize}

To create the edges of the conflict graph, we look to see if the different amino acids are linked in the same way for the two proteins (if, for a given link, the two amino acids linked are the same). Once the conflict graph has been constructed, the similarity between the two molecules is calculated by transforming the conflict graph into an optimisation problem. The proteins can then be sorted according to similarity, from the highest to the lowest with the protein test.

%========================================================================================
\subsection{Construction of conflict graph}
%========================================================================================
The aim is to progressively build the conflict graph for two proteins. The steps involved in constructing this graph are as follows:

\begin{itemize}
\item At the start, the conflict graph is initialized to an empty graph.
\item The algorithm will visit all the amino acids of the two proteins, i.e., the vertices of the graphs represent the proteins.
\item We test whether the two amino acids are more or less similar, i.e., whether the two amino acids have at least one property in common (symbol and/or degree).
\item If this is the case, we will compare the edges linked to these amino acids, or more precisely, we will compare all the vertices adjacent to the current vertices in the same way.
\item A weighted similarity measure is calculated which depends on the two similarities values found: the one given by the vertex comparison and the one given by the edge comparison.
\item  If this measurement value is above a threshold, a certain weight is added according to the similarity of the edges.
\item We add the pair of vertices with their respective weights to the conflict graph as a new vertex.
\item  At this stage, we have a graph with vertices, but no edges. So we're going to go through the vertices of the conflict graph to connect them.
\item For each pair of vertices in the conflict graph, we first check whether an edge is needed and whether one already exists, i.e., whether it is feasible.
\item  Feasibility here means that the amino acids belonging to the first protein are linked in the same way as the amino acids belonging to the second protein.
\item If all the conditions are met, we calculate a weight for the edge and add an edge to the two vertices being compared.
\end{itemize}

\begin{figure}[h!]
\centering
\includegraphics[width=0.99\textwidth]{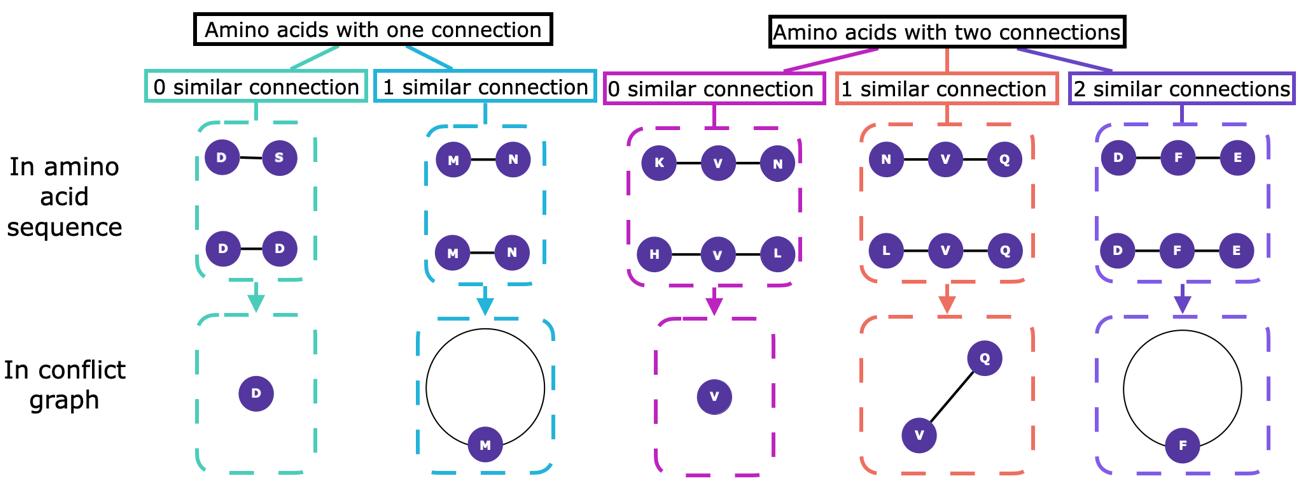}
\caption{Representation of connections between amino acids in the protein sequence and in the conflict graph}
\label{Conflict_graph}
\end{figure}

The constructed conflict graph, denoted $G_c$, is composed of a set of vertices, denoted $V_c$, and a set of edges, denoted $E_c$. The representation of the protein's properties and the link with the amino-acids sequence are represented in Figure \ref{Conflict_graph}.

The resulting conflict graph is an undirected weighted graph.

%========================================================================================
\subsection{Conflict matrix}
%========================================================================================
Now that the conflict graph has been constructed, one of the most common representations of a graph for computing is the adjacency matrix, noted here $A$. In this adjacency matrix, the rows and columns represent each vertices in the graph and each coefficient $A_{i,j}$ is the weight of the edge that connects the vertices $i$ and $j$ or the weight of the vertex if $i$ and $j$ represent the same vertex. Like the conflict graph is an undirected graph, the adjacency matrix is symmetric. It is defined as, 

\begin{equation}
A_{i,j} = \begin{cases} \tilde{w}_i  \mbox{ if } i=j \mbox{ for } i \in V_c \\ w_{ij} \mbox{ for } (i,j) \in E_c
\end{cases}
\end{equation}

where $\tilde{w}_i$ is the weight of the vertex $i$ according to the similarity of the edges and $w_{ij}$ represent the weight of the edge $(i,j)$ . This matrix, which we will  also refer to as the scoring matrix in line with a bioinformatics algorithm (Needleman-Wunsch and Smith-Waterman), will use for the combinatorial optimisation problem to compute the similarity between proteins. 

%========================================================================================
\subsection{Parameters and Similarity rate}
%========================================================================================

For the algorithm with the conflict graph, the first step is to optimise the four parameters:

\begin{itemize}
\item $W_{sim}$: Weight for the similarity between two vertices in the conflict graph
\item $Min_{sim}$: Threshold for the similarity measure between two vertices
\item $W_{edges}$: Weight for the similarity between two edges of the conflict graph
\item $\delta$: Weight of importance between vertex and edge measurements  
\end{itemize}

Like the quantum versions of bioinformatics algorithm, here we are not going to calculate the similarity rate explicitly but the energy rate between the two proteins. First, we must have a reference ration, noted as $\tilde{\Xi}_{ref}$. For this, we calculate the energy ratio between the test protein (with $N_t$ amino acids) and itself. Secondly, we compute the energy ratio between the test protein and the $i^{th}$ protein (with $N_i$ amino acids) of the database (Uniprot), noted as $\tilde{\Xi}_{i}$. This ratio or score, for the Conflict graph algorithm, is defined by:

\begin{equation}
\tilde{\Xi}_{i} = \delta \max \left( \frac{\vert V^{cg}_1 \vert}{\vert V_1 \vert}, \frac{\vert V^{cg}_2 \vert}{\vert V_2 \vert} \right) + (1 - \delta)  \min \left( \frac{\vert V^{cg}_1 \vert}{\vert V_1 \vert}, \frac{\vert V^{cg}_2 \vert}{\vert V_2 \vert} \right)
\end{equation}

\noindent with $\vert V^{cg}_1 \vert$ and $\vert V^{cg}_2 \vert$ the number of unique vertices of the original graphs of the first protein and the second protein in the independent set of the conflict graph, $\vert V_1 \vert$ and $\vert V_2 \vert$ the number of vertices of the original graphs of the first protein and the second protein. The value $\vert V^{cg}_1 \vert$ and $\vert V^{cg}_2 \vert$ are dependant of other parameters ($W_{sim}$, $Min_{sim}$ and $W_{edges}$). The rate of similarity between the two proteins, noted $S_i$, is then defined as follows:

\begin{equation}
S_i = \left( \frac{ \tilde{\Xi}_{i}}{ \tilde{\Xi}_{ref}} \right) \left( \frac{\min (N_t,N_i)}{\max (N_t,N_i)} \right).
\end{equation} 

The aim is to optimise the value of the various parameters in order to achieve the same results as BLASTP. The process will be carried out for all the proteins given in the database.

%========================================================================================
\section{Optimisation problem and resolution}
%========================================================================================

Now we have the scoring matrices (Needleman-Wunsch, Smith-Waterman and Conflict graph algorithm), we search the high rate of similarity between proteins. For this, we use an optimisation problem based on scoring matrices by searching the value of decision variables which maximizes the problem. For the optimisation problem, we take the Quadratic Unconstrained Binary Optimisation (QUBO) problem and for the resolution, we use quantum algorithm: Quantum Approximate Optimisation Algorithm (QAOA).

%========================================================================================
\subsection{Problem model: Quadratic Unconstrained Binary Optimisation (QUBO) problem}
%========================================================================================

The QUBO problem is adapted to our model. Indeed, the vector of decision variable in the problem, noted $x$, will represent either the largest amino-acids sequence between the two proteins being compared (Needleman-Wunsch and Smith-Waterman algorithms), or all the amino acids in common (Conflict graph algorithm). If, an amino acid is considered then $x_i$ takes the value of 1 and 0 if it is not. All decision variables are binary and $x$ belongs to $\lbrace 0, 1 \rbrace^{n_{aa}}$ where $n_{aa}$ is the length of the largest amino-acids sequence for Needleman-Wunsch and Smith-Waterman algorithm, and the number of vertices in the conflict graph for the other algorithm. The general formulation of QUBO is,

\begin{equation}
\begin{cases}
 \mbox{Maximise} & x^T Q x + c^T x  \\
 \mbox{Subject to} & \\ 
 & x \in \lbrace 0,1 \rbrace^{n_{aa}}
\end{cases}
\label{QUBO}
\end{equation}
\noindent where $Q$ is a matrix and $c^T$ is a row vector from $\mathbb{R}^{n_{aa}}$. Given that the decision variables are binary, we assume that $x_i^2 = x_i$, the formulation (\ref{QUBO}) can be rewritten as, 

\begin{equation}
\begin{cases}
 \mbox{Maximise} & x^T \tilde{Q} x   \\
 \mbox{Subject to} & \\ 
 & x \in \lbrace 0,1 \rbrace^{n_{aa}}
\end{cases}
\label{QUBO2}
\end{equation}
\noindent where $\tilde{Q}$ is a matrix with $\tilde{Q}_{i,i} = Q_{i,i} + c_i$. With the scoring matrix, we have,

\begin{equation}
\begin{cases}
 \mbox{Maximise} & x^T M x   \\
 \mbox{Subject to} & \\ 
 & x \in \lbrace 0,1 \rbrace^{n_{aa}}
\end{cases} \Longleftrightarrow 
\begin{cases}
 \mbox{Maximise} & \sum_{i,j} M_{i,j} x_i  x_j    \\
 \mbox{Subject to} & \\ 
 & x_i \in \lbrace 0,1 \rbrace, \forall i
\end{cases}
\label{matrix_QUBO}
\end{equation}

\noindent where $M$ is the scoring matrix. In graph theory, the QUBO problem is defined with a graph, denoted by $G$, then it is composed of a set of vertices, denoted $V$, a set of edges, denoted $E$ and a set of the weights of edges, denoted $W$ \cite{Punnen2022}. The optimisation problem can also be formulated as follows,

\begin{equation}
\begin{cases}
 \mbox{Maximise} &  \sum_{i \in V} \tilde{w}_{i} x_i - \sum_{(i,j) \in E} w_{ij} x_i x_j    \\
 \mbox{Subject to} & \\ 
 &  x_i \in \lbrace 0,1 \rbrace, \forall i \in V \mbox{ for the graph } G = (V,E)
\end{cases}
\label{graph_QUBO}
\end{equation}

The two formulations (\ref{matrix_QUBO}) and (\ref{graph_QUBO}) are equivalent with the adjacency matrix $A$ as the scoring matrix. The weights $\tilde{w}_{i}$ correspond to the diagonal coefficients of the scoring matrix and the weights $w_{ij}$ correspond to the off-diagonal coefficients. In all cases we will use the matrix formulation (\ref{matrix_QUBO}) even if the graph formulation (\ref{graph_QUBO}) would avoid the construction of the scoring matrix for the Conflict graph algorithm.

 Now to solve this problem, we use a quantum method that allows the calculations to be carried out in parallel more quickly thanks to the formalisms of quantum mechanics. In addition, this method is coupled with an optimiser to find the values of the parameters required for the quantum computations. At the end, we find the percentage of similarity between the two proteins.

%========================================================================================
\subsection{Solver: Quantum Approximate Optimisation Algorithm (QAOA)}
%========================================================================================
To solve the QUBO problem, we'll use Quantum Approximate Optimisation Algorithm (QAOA). This method is a quantum algorithm that gives an approximate solutions for combinatorial optimisation problems \cite{Farhi2014}. It is a hybrid algorithm that uses quantum and classical capacities with the aim to reduce resources of the quantum part, i.e., the number of qubits required. The solution of QUBO is equivalent to the ground state of a corresponding Hamiltonian \cite{Moll2018}. The quantum part uses two concepts of the quantum mechanics, the phase of a state (in particular the ground state) and its probabilities. QAOA aims to find feasible solutions to the QUBO problems by minimizing the expected value of two Hamiltonian $H_C$ and $H_M$, that describe the energy of a quantum system and follows the equation of Schrödinger,

\begin{equation}
\frac{\partial}{\partial t} \vert \psi_t \rangle = - \frac{i}{\hbar} \left(\alpha(t) H_C + \left( 1-\alpha(t) \right) H_M \right) \vert \psi_t \rangle
\end{equation}

\noindent $\hbar$ is derived from Planck constant, $\vert \psi_t \rangle$ are states vectors and $\alpha(t)$ is one parameter for interpolation and smoothly decreases from 1 to 0 \cite{Farhi2000}. The Hamiltonian $H_C$ corresponds to the Hamiltonian that encodes the QUBO problem, called the cost Hamiltonian.  The second Hamiltonian $H_M$, commonly referred to as the \textquotedblleft{} driver Hamiltonian \textquotedblright{}, should not commute with the cost Hamiltonian $H_C$, called the mixing Hamiltonian. The step of QAOA are the following:

\begin{enumerate}
\item Prepare the quantum system in superposition state (use the Hadamard gate on each qubit).
\item Define the cost Hamiltonian $H_C$.
\item Define the mixing Hamiltonian $H_M$.
\item Build the unitaries gates $U(\gamma, H_C)$ and $U(\beta, H_M)$ defined by,
\begin{align*}
& U(\gamma, H_C) = \text{e}^{-i \gamma H_C} & U(\beta, H_M) = \text{e}^{-i \beta H_M}
\end{align*}

We can't use directly Hamiltonian matrices because they aren't unitary. Thus, we must use the exponential of that matrices. The unitary gate $U(\gamma, H_C)$ performs a phase rotation on the Z-axis. The phase of each state obtained is proportional to the corresponding eigenvalue. The unitary gate $U(\beta, H_M)$ performs a phase rotation on the X-axis. They transform the phase of each state into amplitude, this means that the states with more phase, or higher eigenvalue, have a greater probability of being measured.  

\item Construct the quantum circuit by applying $p$ times  $U(\gamma, H_C)$ and  $U(\beta, H_M)$ to the superposition state,

\begin{equation}
QAOA (\gamma, \beta) = \text{e}^{-i \gamma_1 H_C}  \text{e}^{-i \beta_1 H_M} \cdots \text{e}^{-i \gamma_p H_C}  \text{e}^{-i \beta_p H_M}.
\end{equation}
\item Measure the state of the quantum system.
\item Change the values of $\gamma_i$ and $\beta_i$ with a classical optimiser until a satisfactory state that will represent the solution of QUBO problem. 
\end{enumerate}

\begin{figure}[h!]
\centering
\includegraphics[width=0.8\textwidth]{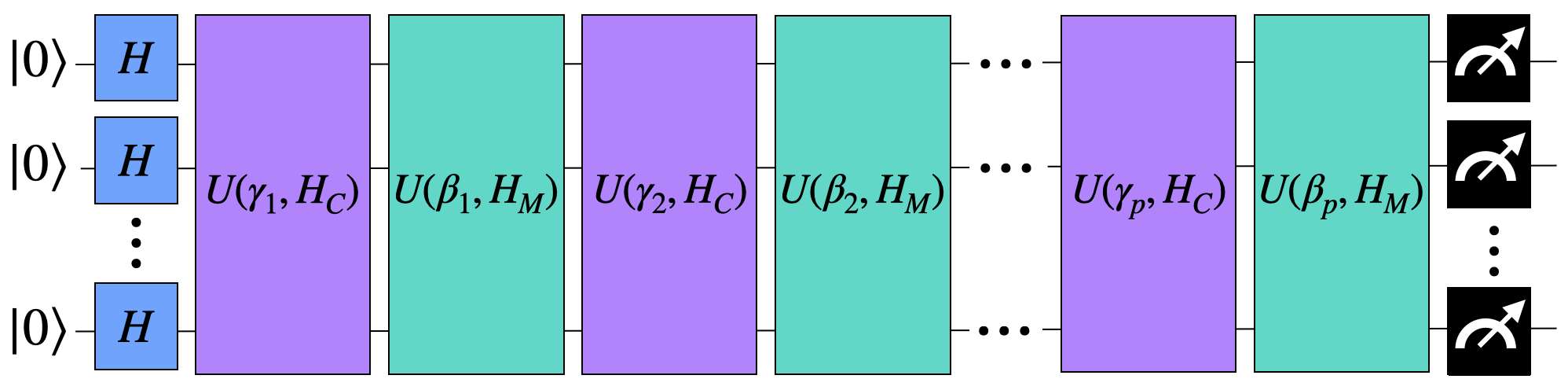}
\caption{Circuit representation for QAOA algorithm}
\label{Qaoa_circ}
\end{figure}

The quantum part of QAOA is represented in Figure \ref{Qaoa_circ}.

%========================================================================================
\section{Numerical results}
%========================================================================================
The results of the various algorithms will be compared with the results of our Protein-BLAST reference. The latter gives us several pieces of information, such as the percentage of the test sequence that overlaps that of the current protein in the database (called Query cover), the percentage of identical base pairs between the test sequence and that of the protein in the database (called Identity) and the alignment score (called Score). In the first time, the result will be compared for human elafin and in a second time for a protein test created with the quantum protein generator.

\subsection{Results for Elafin}

As both bioinformatics algorithms (Needleman-Wunsch and Smith-Waterman) are alignment algorithms, only the alignment score for BLASTP will be considered. In addition, the results of BLASTP will be compared with those of the quantum algorithms, its alignment scores will be normalised and we will focus on the four proteins closest to human elafin,  ELAF\_HUMAN (Table \ref{Blastp}), i.e., those who have a score above 80. As with the other results, the BLASTP results will be normalised to the size of the sequences relative to that of the human elafin, giving us the similarity.

\begin{table}[h!]
\centering
\begin{tabular}{|c|c|c|c|c|c|c|}
\hline
\rowcolor[HTML]{A490DD} 
Protein name & \begin{tabular}[c]{@{}c@{}}Sequence \\ length\end{tabular} & \begin{tabular}[c]{@{}c@{}}Query \\ cover\end{tabular} & Score & Identity & \begin{tabular}[c]{@{}c@{}}Score \\ in percentage \end{tabular} & Similarity \\ \hline
\rowcolor[HTML]{EBE9F5} 
ELAF\_HUMAN & 117 & 1.000 & 231 & 1.000 & 1.000 & 1.000  \\ \hline
\rowcolor[HTML]{EBE9F5} 
ELAF\_PIG  & 167 & 1.000 & 115 & 0.485 & 0.498 & 0.349  \\ \hline
\rowcolor[HTML]{EBE9F5} 
WAP3\_PIG & 144 & 1.000 & 93.2 & 0.441 & 0.403 & 0.329 \\ \hline
\rowcolor[HTML]{EBE9F5} 
SPAI\_PIG  & 187 & 1.000 & 99.4 & 0.380 & 0.430 & 0.269 \\ \hline
\end{tabular}
\caption{BLASTP results}
\label{Blastp}
\end{table}

We have used Python 3.9 to implement our algorithms, which run on an Intel Core i7 with 2.9 GHz and 16 GB of RAM with Microsoft Windows 11 OS. The quantum part is implemented with the Qiskit library and for validating the result, programs are launched on IBM quantum processors.

\subsubsection{Results for algorithm with conflict graph}

The aim is to optimize the parameters $\left( W_{sim}, Min_{sim}, W_{edges} , \delta \right)$, to see the influence of these parameters on the repartition of similarity measures and see the four proteins that are closest to human elafin. For the reference values of the parameters, we took the same values as in \cite{Jimenez21}, i.e., $\left( W_{sim}, Min_{sim}, W_{edges} , \delta \right) = \left( 0.30, 0.75,1.00,0.50 \right)$. We will focus on just one parameter each time, with the others having their reference values. As a reminder,

\begin{itemize}
\item $W_{sim}$: Weight for the similarity between two vertices on the conflict graph
\item $Min_{sim}$: Threshold for the similarity measure between two vertices
\item $W_{edges}$: Weight for the similarity between two edges on the conflict graph
\item $\delta$: Weight of importance between vertex and edge measurements  
\end{itemize}

For the influence of $W_{sim}$, the parameter takes as values $0.30$, $0.50$, $0.75$ and $1.00$. Let's take a closer look at the similarity podiums (Table \ref{Wsimpodium}).

\begin{table}[h!]
\centering
\begin{subtable}{0.49\textwidth}
\centering
\resizebox{\textwidth}{!}{
\begin{tabular}{|c|c|c|c|}
\hline
\rowcolor[HTML]{A490DD} 
Protein name & \begin{tabular}[c]{@{}c@{}}Sequence \\ length\end{tabular} & Score & Similarity \\ \hline
\rowcolor[HTML]{EBE9F5} 
ELAF\_HUMAN  & 117 & 1.000 & 1.000 \\ \hline
\rowcolor[HTML]{EBE9F5} 
ELAF\_PIG  & 167 & 0.428 & 0.300 \\ \hline
\rowcolor[HTML]{EBE9F5} 
WAP3\_PIG & 144 & 0.326 & 0.265 \\ \hline
\rowcolor[HTML]{EBE9F5} 
PER\_DROOR  & 114 & 0.253 & 0.247 \\ \hline
\end{tabular}}
\caption{$W_{sim} = 0.30$ and $W_{sim} = 1.00$}
\label{Wsim03}
\end{subtable}
\hfill
\begin{subtable}{0.49\textwidth}
\centering
\resizebox{\textwidth}{!}{
\begin{tabular}{|c|c|c|c|}
\hline
\rowcolor[HTML]{A490DD} 
Protein name & \begin{tabular}[c]{@{}c@{}}Sequence \\ length\end{tabular} & Score & Similarity \\ \hline
\rowcolor[HTML]{EBE9F5} 
ELAF\_HUMAN  & 117 & 1.000 & 1.000 \\ \hline
\rowcolor[HTML]{EBE9F5} 
NLTP3\_HORVU  & 118 & 0.498 & 0.494 \\ \hline
\rowcolor[HTML]{EBE9F5} 
WFD12\_MACMU & 111 & 0.500 & 0.474 \\ \hline
\rowcolor[HTML]{EBE9F5} 
WFD12\_PAPAN & 111 & 0.500 & 0.474 \\ \hline
\end{tabular}}
\caption{$W_{sim} = 0.50$ and $W_{sim} = 0.75$}
\label{Wsim05}
\end{subtable}
\caption{Podium for $W_{sim}$ parameter}
\label{Wsimpodium}
\end{table}

We have for $W_{sim} = 0.30$ and $W_{sim} = 1.00$ (Table \ref{Wsim03}) the same podium but also when $W_{sim} = 0.50$ and $W_{sim} = 0.75$ (Table \ref{Wsim05}). We note that for $W_{sim} = 0.30$ and $W_{sim} = 1.00$ (Table \ref{Wsim03}), we find the only three proteins closest to human elafin (ELAF\_HUMAN, ELAF\_PIG, WAP3\_PIG) and SPAI\_PIG is missing compared to BLASTP. Whereas for $W_{sim} = 0.50$ and $W_{sim} = 0.75$ (Table \ref{Wsim05}), only ELAF\_HUMAN remains in the podium. The similarity found is lower than the BLASTP results. We have a difference of 2.0 \% for ELAF\_PIG and 3.8 \% for WAP3\_PIG.  For $W_{sim}$, we keep the same value as \cite{Jimenez21}, i.e., $W_{sim} = 0.30$ and change other parameters to find SPAI\_PIG in the podium. 

 Now, for the influence of $Min_{sim}$, the parameter takes as values $0.50$, $0.75$ and $1.00$. 

\begin{table}[h!]
\centering
\begin{subtable}{0.49\textwidth}
\centering
\resizebox{\textwidth}{!}{
\begin{tabular}{|c|c|c|c|}
\hline
\rowcolor[HTML]{A490DD} 
Protein name & \begin{tabular}[c]{@{}c@{}}Sequence \\ length\end{tabular} & Score & Similarity \\ \hline
\rowcolor[HTML]{EBE9F5} 
ELAF\_HUMAN & 117 & 1.000 & 1.000 \\ \hline
\rowcolor[HTML]{EBE9F5} 
GUC2A\_HUMAN & 115 & 0.866 & 0.852 \\ \hline
\rowcolor[HTML]{EBE9F5} 
NLTP\_LACSA & 117 & 0.846 & 0.846 \\ \hline
\rowcolor[HTML]{EBE9F5} 
HYPA\_DESOH  & 115 & 0.825 & 0.818 \\ \hline
\end{tabular}}
\caption{$Min_{sim} = 0.25$}
\label{Msimtable025}
\end{subtable}
\hfill
\begin{subtable}{0.49\textwidth}
\centering
\resizebox{\textwidth}{!}{
\begin{tabular}{|c|c|c|c|}
\hline
\rowcolor[HTML]{A490DD} 
Protein name & \begin{tabular}[c]{@{}c@{}}Sequence \\ length\end{tabular} & Score & Similarity \\ \hline
\rowcolor[HTML]{EBE9F5} 
ELAF\_HUMAN & 117 & 1.000 & 1.000 \\ \hline
\rowcolor[HTML]{EBE9F5} 
NLTP3\_HORVU & 118 & 0.566 & 0.561 \\ \hline
\rowcolor[HTML]{EBE9F5} 
PSMG3\_HUMAN & 122 & 0.544 & 0.522 \\ \hline
\rowcolor[HTML]{EBE9F5} 
WFD12\_PAPAN  & 111 & 0.544 & 0.516 \\ \hline
\end{tabular}}
\caption{$Min_{sim} = 0.50$}
\label{Msimtable05}
\end{subtable}
\hfill
\begin{subtable}{0.49\textwidth}
\centering
\resizebox{\textwidth}{!}{
\begin{tabular}{|c|c|c|c|}
\hline
\rowcolor[HTML]{A490DD} 
Protein name & \begin{tabular}[c]{@{}c@{}}Sequence \\ length\end{tabular} & Score & Similarity \\ \hline
\rowcolor[HTML]{EBE9F5} 
ELAF\_HUMAN  & 117 & 1.000 & 1.000 \\ \hline
\rowcolor[HTML]{EBE9F5} 
ELAF\_PIG  & 167 & 0.428 & 0.300 \\ \hline
\rowcolor[HTML]{EBE9F5} 
WAP3\_PIG & 144 & 0.326 & 0.265 \\ \hline
\rowcolor[HTML]{EBE9F5} 
PER\_DROOR & 114 & 0.253 & 0.247 \\ \hline
\end{tabular}}
\caption{$Min_{sim} = 0.75$ and $Min_{sim} = 1.00$}
\label{Msimtable1}
\end{subtable}
\caption{Podium for $Min_{sim}$ parameter}
\label{Msimtable}
\end{table}

The podium is the same for $Min_{sim} = 0.75$ and $Min_{sim} = 1.00$ (Table \ref{Msimtable1}). For $Min_{sim} = 0.25$ and $Min_{sim} = 0.50$, we have a completely different podium to that of BLASTP and no reference protein except ELAF\_HUMAN can be found in it (Table \ref{Msimtable05}). The threshold must have a value greater than 0.5 to obtain results close to BLASTP. The differences for ELAF\_PIG and WAP3\_PIG compared to the reference are the same as the previous parameter. For $Min_{sim}$, we keep the same value as \cite{Jimenez21}, i.e., $Min_{sim} = 0.75$. For the influence of $W_{edges}$, the parameter takes as values $0.00$, $0.25$, $0.50$, $0.75$ and $1.00$. 
 
 \begin{table}[h!]
\centering
\begin{tabular}{|c|c|c|c|}
\hline
\rowcolor[HTML]{A490DD} 
Protein name & \begin{tabular}[c]{@{}c@{}}Sequence \\ length\end{tabular} & Score & Similarity \\ \hline
\rowcolor[HTML]{EBE9F5} 
ELAF\_HUMAN  & 117 & 1.000 & 1.000 \\ \hline
\rowcolor[HTML]{EBE9F5} 
ELAF\_PIG  & 167 & 0.428 & 0.300 \\ \hline
\rowcolor[HTML]{EBE9F5} 
WAP3\_PIG  & 144 & 0.326 & 0.265 \\ \hline
\rowcolor[HTML]{EBE9F5} 
PER\_DROOR & 114 & 0.253 & 0.247 \\ \hline
\end{tabular}
\caption{Podium for $W_{edge}$ parameter}
\label{Wedge}
\end{table}

 We have the same protein similarity podium for all values of $W_{edges}$ (Table \ref{Wedge}). Already in \cite{Jimenez21}, the parameter $W_{edges}$ had very little influence on the results. We keep the reference value, i.e., $W_{edge} = 1.00$. And finally for the influence of $\delta$, the parameter takes as values $0.00$, $0.25$, $0.50$, $0.75$ and $1.00$.

\begin{table}[h!]
\centering
\begin{subtable}{0.49\textwidth}
\centering
\resizebox{\textwidth}{!}{
\begin{tabular}{|c|c|c|c|}
\hline
\rowcolor[HTML]{A490DD} 
Protein name & \begin{tabular}[c]{@{}c@{}}Sequence \\ length\end{tabular} & Score & Similarity \\ \hline
\rowcolor[HTML]{EBE9F5} 
ELAF\_HUMAN  & 117 & 1.000 & 1.000 \\ \hline
\rowcolor[HTML]{EBE9F5} 
ELAF\_PIG  & 167 & 0.395 & 0.277 \\ \hline
\rowcolor[HTML]{EBE9F5} 
WAP3\_PIG & 144 & 0.319 & 0.260 \\ \hline
\rowcolor[HTML]{EBE9F5} 
SPAI\_PIG  & 187 & 0.368 & 0.230 \\ \hline
\end{tabular}}
\caption{$\delta = 0.00$}
\label{delta0}
\end{subtable}
\hfill
\begin{subtable}{0.49\textwidth}
\centering
\resizebox{\textwidth}{!}{
\begin{tabular}{|c|c|c|c|}
\hline
\rowcolor[HTML]{A490DD} 
Protein name & \begin{tabular}[c]{@{}c@{}}Sequence \\ length\end{tabular} & Score & Similarity \\ \hline
\rowcolor[HTML]{EBE9F5} 
ELAF\_HUMAN  & 117 & 1.000 & 1.000 \\ \hline
\rowcolor[HTML]{EBE9F5} 
ELAF\_PIG  & 167 & 0.412 & 0.289 \\ \hline
\rowcolor[HTML]{EBE9F5} 
WAP3\_PIG  & 144 & 0.323 & 0.262 \\ \hline
\rowcolor[HTML]{EBE9F5} 
SPAI\_PIG & 187 & 0.372 & 0.233 \\ \hline
\end{tabular}}
\caption{$\delta = 0.25$}
\label{delta025}
\end{subtable}

\begin{subtable}{0.49\textwidth}
\centering
\resizebox{\textwidth}{!}{
\begin{tabular}{|c|c|c|c|}
\hline
\rowcolor[HTML]{A490DD} 
Protein name & \begin{tabular}[c]{@{}c@{}}Sequence \\ length\end{tabular} & Score & Similarity \\ \hline
\rowcolor[HTML]{EBE9F5} 
ELAF\_HUMAN  & 117 & 1.000 & 1.000 \\ \hline
\rowcolor[HTML]{EBE9F5} 
ELAF\_PIG  & 167 & 0.428 & 0.300 \\ \hline
\rowcolor[HTML]{EBE9F5} 
WAP3\_PIG & 144 & 0.326 & 0.265 \\ \hline
\rowcolor[HTML]{EBE9F5} 
PER\_DROOR & 114 & 0.253 & 0.247 \\ \hline
\end{tabular}}
\caption{$\delta = 0.50$}
\label{delta05}
\end{subtable}
\hfill
\begin{subtable}{0.49\textwidth}
\centering
\resizebox{\textwidth}{!}{
\begin{tabular}{|c|c|c|c|}
\hline
\rowcolor[HTML]{A490DD} 
Protein name & \begin{tabular}[c]{@{}c@{}}Sequence \\ length\end{tabular} & Score & Similarity \\ \hline
\rowcolor[HTML]{EBE9F5} 
ELAF\_HUMAN  & 117 & 1.000 & 1.000 \\ \hline
\rowcolor[HTML]{EBE9F5} 
PER\_DROOR  & 114 & 0.333 & 0.324 \\ \hline
\rowcolor[HTML]{EBE9F5} 
ELAF\_PIG  & 167 & 0.445 & 0.312 \\ \hline
\rowcolor[HTML]{EBE9F5} 
PER\_DROAN & 125 & 0.330 & 0.309 \\ \hline
\end{tabular}}
\caption{$\delta = 0.75$}
\label{delta075}
\end{subtable}
\begin{subtable}{0.49\textwidth}
\centering
\resizebox{\textwidth}{!}{
\begin{tabular}{|c|c|c|c|}
\hline
\rowcolor[HTML]{A490DD} 
Protein name & \begin{tabular}[c]{@{}c@{}}Sequence \\ length\end{tabular} & Score & Similarity \\ \hline
\rowcolor[HTML]{EBE9F5} 
ELAF\_HUMAN  & 117 & 1.000 & 1.000 \\ \hline
\rowcolor[HTML]{EBE9F5} 
PER\_DROOR & 114 & 0.412 & 0.402 \\ \hline
\rowcolor[HTML]{EBE9F5} 
Y8587\_DICDI & 69 & 0.652 & 0.385 \\ \hline
\rowcolor[HTML]{EBE9F5} 
PER\_DROAN & 125 & 0.400 & 0.374 \\ \hline
\end{tabular}}
\caption{$\delta = 1.00$}
\label{delta1}
\end{subtable}
\caption{Podium for $\delta$ parameter}
\label{deltapodium}
\end{table}

For the podium, the $\delta$ parameter has a huge influence on the latter (Table \ref{deltapodium}) and increase the similarity measure of all proteins on the database. However, for $\delta > 0.5$, the podium obtained no longer corresponds to the results of BLASTP (Table \ref{delta075} and Table \ref{delta1}), i.e., especially the four proteins (ELAF\_HUMAN, ELAF\_PIG, WAP3\_PIG and SPAI\_PIG). For $\delta = 0.5$, the protein SPAI\_PIG is always missing. As for $\delta = 0$ and $\delta = 0.25$, the podium is the same and we have the same four proteins compared to BLASTP. The only difference is noted for the similarity measure. These latter is higher for $\delta = 0.25$. For ELAF\_PIG, we have a difference of 4.3 \% for $\delta = 0.00$ and 3.0 \% for $\delta = 0.25$. For WAP3\_PIG, the difference is 4.4 \% for $\delta = 0.00$ and 1.5 \% for $\delta = 0.25$. And finally for SPAI\_PIG, we have a difference of 2.1 \% for $\delta = 0.00$ and 1.8 \% for $\delta = 0.25$. Given the results for $\delta$, this time we will use $\delta = 0.25$ instead of using the same value as \cite{Jimenez21}.

\subsubsection{Results for Classical and Quantum Needleman-Wunsch algorithm}

For the Needleman-Wunsch algorithms (Classical and Quantum), we are going to see the influence of the gap penalty parameter with $GP_1 = \lbrace 1,2,4 \rbrace$. As with the Conflict graph algorithm, we are interested in the podiums found.

 For the classical podiums (Table \ref{CNMpodium}), we found the same four proteins compared to BLASTP and only for $GP_1 = 4$ their order change. We also see that as the gap penalty value increases, the similarity rate decreases. For ELAF\_PIG, we have a difference of 77.9 \% for $GP_1 = 1$ and 34.5 \% for $GP_1 = 2$. For WAP3\_PIG, the difference is 74.5 \% for $GP_1 = 1$ and 34.1 \% for $GP_1 = 2$. And finally for SPAI\_PIG, we have a difference of 156.9 \% for $GP_1 = 1$ and 58.1 \% for $GP_1 = 2$. For classical Needleman-Wunsch algorithm, the gap penalty $GP_1 = 2$ gives the results closest to BLASTP giving priority to its order. But for the difference, the gap penalty $GP_1 = 4$ gives the results closest to the reference. For $GP_1 = 4$, we have a difference of 8.4 \% for ELAF\_PIG, 1.2 \% for WAP3\_PIG and 2.9 \% for SPAI\_PIG.

\begin{table}[h!]
\centering
\begin{subtable}{0.49\textwidth}
\centering
\resizebox{\textwidth}{!}{
\begin{tabular}{|c|c|c|c|}
\hline
\rowcolor[HTML]{A490DD} 
Protein name & \begin{tabular}[c]{@{}c@{}}Sequence \\ length\end{tabular} & Score & Similarity \\ \hline
\rowcolor[HTML]{EBE9F5} 
ELAF\_HUMAN  & 117 & 1.000 & 1.000 \\ \hline
\rowcolor[HTML]{EBE9F5} 
ELAF\_PIG  & 167 & 0.938 & 0.657 \\ \hline
\rowcolor[HTML]{EBE9F5} 
WAP3\_PIG & 144 & 0.755 & 0.613 \\ \hline
\rowcolor[HTML]{EBE9F5} 
SPAI\_PIG & 187 & 0.968 & 0.606 \\ \hline
\end{tabular}}
\caption{Gap penalty $= 1$}
\label{CNMP1}
\end{subtable}
\hfill
\begin{subtable}{0.49\textwidth}
\centering
\resizebox{\textwidth}{!}{
\begin{tabular}{|c|c|c|c|}
\hline
\rowcolor[HTML]{A490DD} 
Protein name & \begin{tabular}[c]{@{}c@{}}Sequence \\ length\end{tabular} & Score & Similarity \\ \hline
\rowcolor[HTML]{EBE9F5} 
ELAF\_HUMAN  & 117 & 1.000 & 1.000 \\ \hline
\rowcolor[HTML]{EBE9F5} 
ELAF\_PIG  & 167 & 0.791 & 0.554 \\ \hline
\rowcolor[HTML]{EBE9F5} 
WAP3\_PIG  & 144 & 0.642 & 0.521 \\ \hline
\rowcolor[HTML]{EBE9F5} 
SPAI\_PIG & 187 & 0.758 & 0.474 \\ \hline
\end{tabular}}
\caption{Gap penalty $= 2$}
\label{CNMP2}
\end{subtable}
\begin{subtable}{0.49\textwidth}
\centering
\resizebox{\textwidth}{!}{
\begin{tabular}{|c|c|c|c|}
\hline
\rowcolor[HTML]{A490DD} 
Protein name & \begin{tabular}[c]{@{}c@{}}Sequence \\ length\end{tabular} & Score & Similarity \\ \hline
\rowcolor[HTML]{EBE9F5} 
ELAF\_HUMAN & 117 & 1.000 & 1.000 \\ \hline
\rowcolor[HTML]{EBE9F5} 
WAP3\_PIG & 144 & 0.453 & 0.368 \\ \hline
\rowcolor[HTML]{EBE9F5} 
ELAF\_PIG  & 167 & 0.457 & 0.320 \\ \hline
\rowcolor[HTML]{EBE9F5} 
SPAI\_PIG & 187 & 0.359 & 0.224 \\ \hline
\end{tabular}}
\caption{Gap penalty $= 4$}
\label{CNMP4}
\end{subtable}
\caption{Podium for Classical Needleman-Wunsch with various gap penalties}
\label{CNMpodium}
\end{table}

For the quantum podiums (Table \ref{QNMpodium}), however we found the same four proteins compared to BLASTP only for $GP_1 = 4$. We have a difference of 2.2 \% for ELAF\_PIG, 2.6 \% for WAP3\_PIG and 3.5 \% for SPAI\_PIG.

\begin{table}[h!]
\centering
\begin{subtable}{0.49\textwidth}
\centering
\resizebox{\textwidth}{!}{
\begin{tabular}{|c|c|c|c|}
\hline
\rowcolor[HTML]{A490DD} 
Protein name & \begin{tabular}[c]{@{}c@{}}Sequence \\ length\end{tabular} & Score & Similarity \\ \hline
\rowcolor[HTML]{EBE9F5} 
ELAF\_HUMAN & 117 & 2167604 & 1.000 \\ \hline
\rowcolor[HTML]{EBE9F5} 
ELAF\_PIG  & 167 & 2142114 & 0.692 \\ \hline
\rowcolor[HTML]{EBE9F5} 
SPAI\_PIG & 187 & 2229906 & 0.644 \\ \hline
\rowcolor[HTML]{EBE9F5} 
WAP3\_PIG & 144 & 1582054 & 0.593 \\ \hline
\end{tabular}}
\caption{Gap penalty $= 1$}
\label{QNMP1}
\end{subtable}
\hfill
\begin{subtable}{0.49\textwidth}
\centering
\resizebox{\textwidth}{!}{
\begin{tabular}{|c|c|c|c|}
\hline
\rowcolor[HTML]{A490DD} 
Protein name & \begin{tabular}[c]{@{}c@{}}Sequence \\ length\end{tabular} & Score & Similarity \\ \hline
\rowcolor[HTML]{EBE9F5} 
ELAF\_HUMAN  & 117 & 1878542 & 1.000 \\ \hline
\rowcolor[HTML]{EBE9F5} 
ELAF\_PIG  & 167 & 1529215 & 0.570 \\ \hline
\rowcolor[HTML]{EBE9F5} 
SPAI\_PIG & 187 & 1566335 & 0.522 \\ \hline
\rowcolor[HTML]{EBE9F5} 
WAP3\_PIG & 144 & 1127360 & 0.488 \\ \hline
\end{tabular}}
\caption{Gap penalty $= 2$}
\label{QNMP2}
\end{subtable}
\begin{subtable}{0.49\textwidth}
\centering
\resizebox{\textwidth}{!}{
\begin{tabular}{|c|c|c|c|}
\hline
\rowcolor[HTML]{A490DD} 
Protein name & \begin{tabular}[c]{@{}c@{}}Sequence \\ length\end{tabular} & Score & Similarity \\ \hline
\rowcolor[HTML]{EBE9F5} 
ELAF\_HUMAN  & 117 & 1359755 & 1.000 \\ \hline
\rowcolor[HTML]{EBE9F5} 
ELAF\_PIG   & 167 & 575566 & 0.297 \\ \hline
\rowcolor[HTML]{EBE9F5} 
WAP3\_PIG  & 144 & 462188 & 0.276 \\ \hline
\rowcolor[HTML]{EBE9F5} 
SPAI\_PIG & 187 & 475160 & 0.219 \\ \hline
\end{tabular}}
\caption{Gap penalty $= 4$}
\label{QNMP4}
\end{subtable}
\caption{Podium for Quantum Needleman-Wunsch with various gap penalties}
\label{QNMpodium}
\end{table}

During the simulations, we realised that the Quantum Needleman-Wunsch algorithm takes a very long time (or even does not converge) for certain proteins which are sometimes very small proteins of less than 50 amino acids. The characteristic of these proteins is that the spectrum of the matrix (dotplot and Blosum matrices taken into consideration) has a globally symmetrical spectrum (if the eigenvalue $\lambda$ belongs to the spectrum then $-\lambda$ also belongs). If the spectrum is symmetric, then the resulting linear application is stable, which implies that the solution Hilbert space is also stable (linear algebra) or globally invariant (geometry). If this space is globally invariant, then the qubits vary very little and QAOA has difficulty in finding the fundamental state of the quantum system, resulting in a very slow resolution time. For other proteins or with the Quantum Smith-Waterman algorithm, the spectrum of the matrix is not symmetrical, which results in relatively low resolution times.

\subsubsection{Results for Classical and Quantum Smith-Waterman algorithm}

For the Smith-Waterman algorithms (Classical and Quantum), we are going to see the influence of the gap penalty parameter with $GP_1 = \lbrace 1,2,4 \rbrace$. As with the other algorithm, we are interested in the podiums found.

 For the classical podiums (Table \ref{CSWpodium}), we found the same four proteins compared to BLASTP and only for $GP_1 = 4$ their order change. We also see that as the gap penalty value increases, the similarity rate decreases. For ELAF\_PIG, we have a difference of 77.9 \% for $GP_1 = 1$ and 34.5 \% for $GP_1 = 2$. For WAP3\_PIG, the difference is 74.5 \% for $GP_1 = 1$ and 34.1 \% for $GP_1 = 2$. And finally for SPAI\_PIG, we have a difference of 156.9 \% for $GP_1 = 1$ and 58.1 \% for $GP_1 = 2$. For the classical Smith-Waterman algorithm, the gap penalty $GP_1 = 2$ gives the results closest to BLASTP giving priority to its order. But for the difference, the gap penalty $GP_1 = 4$ gives the results closest to the reference. For $GP_1 = 4$, we have a difference of 8.4 \% for ELAF\_PIG, 1.2 \% for WAP3\_PIG and 2.9 \% for SPAI\_PIG.

\begin{table}[h!]
\centering
\begin{subtable}{0.49\textwidth}
\centering
\resizebox{\textwidth}{!}{
\begin{tabular}{|c|c|c|c|}
\hline
\rowcolor[HTML]{A490DD} 
Protein name & \begin{tabular}[c]{@{}c@{}}Sequence \\ length\end{tabular} & Score & Similarity \\ \hline
\rowcolor[HTML]{EBE9F5} 
ELAF\_HUMAN  & 117 & 1.000 & 1.000 \\ \hline
\rowcolor[HTML]{EBE9F5} 
ELAF\_PIG  & 167 & 0.938 & 0.657 \\ \hline
\rowcolor[HTML]{EBE9F5} 
WAP3\_PIG  & 144 & 0.755 & 0.613 \\ \hline
\rowcolor[HTML]{EBE9F5} 
SPAI\_PIG & 187 & 0.968 & 0.606 \\ \hline
\end{tabular}}
\caption{Gap penalty $= 1$}
\label{CSWP1}
\end{subtable}
\hfill
\begin{subtable}{0.49\textwidth}
\centering
\resizebox{\textwidth}{!}{
\begin{tabular}{|c|c|c|c|}
\hline
\rowcolor[HTML]{A490DD} 
Protein name & \begin{tabular}[c]{@{}c@{}}Sequence \\ length\end{tabular} & Score & Similarity \\ \hline
\rowcolor[HTML]{EBE9F5} 
ELAF\_HUMAN  & 117 & 1.000 & 1.000 \\ \hline
\rowcolor[HTML]{EBE9F5} 
ELAF\_PIG  & 167 & 0.791 & 0.554 \\ \hline
\rowcolor[HTML]{EBE9F5} 
WAP3\_PIG & 144 & 0.642 & 0.521 \\ \hline
\rowcolor[HTML]{EBE9F5} 
SPAI\_PIG & 187 & 0.758 & 0.474 \\ \hline
\end{tabular}}
\caption{Gap penalty $= 2$}
\label{CSWP2}
\end{subtable}
\begin{subtable}{0.49\textwidth}
\centering
\resizebox{\textwidth}{!}{
\begin{tabular}{|c|c|c|c|}
\hline
\rowcolor[HTML]{A490DD} 
Protein name & \begin{tabular}[c]{@{}c@{}}Sequence \\ length\end{tabular} & Score & Similarity \\ \hline
\rowcolor[HTML]{EBE9F5} 
ELAF\_HUMAN  & 117 & 1.000 & 1.000 \\ \hline
\rowcolor[HTML]{EBE9F5} 
WAP3\_PIG & 144 & 0.360 & 0.293 \\ \hline
\rowcolor[HTML]{EBE9F5} 
ELAF\_PIG & 167 & 0.354 & 0.248 \\ \hline
\rowcolor[HTML]{EBE9F5} 
SPAI\_PIG & 187 & 0.357 & 0.223 \\ \hline
\end{tabular}}
\caption{Gap penalty $= 4$}
\label{CSWP4}
\end{subtable}
\caption{Podium for Classical Smith-Waterman with various gap penalties}
\label{CSWpodium}
\end{table}

For the quantum podiums (Table \ref{QSWpodium}), however we found the same four proteins compared to BLASTP only their order change. For $GP_1 = 4$, we have a difference of 5.1 \% for ELAF\_PIG, 2.2 \% for WAP3\_PIG and 38.5 \% for SPAI\_PIG.

\begin{table}[h!]
\centering
\begin{subtable}{0.49\textwidth}
\centering
\resizebox{\textwidth}{!}{
\begin{tabular}{|c|c|c|c|}
\hline
\rowcolor[HTML]{A490DD} 
Protein name & \begin{tabular}[c]{@{}c@{}}Sequence \\ length\end{tabular} & Score & Similarity \\ \hline
\rowcolor[HTML]{EBE9F5} 
ELAF\_HUMAN  & 117 & 2197821 & 1.000 \\ \hline
\rowcolor[HTML]{EBE9F5} 
ELAF\_PIG  & 167 & 2217077 & 0.707 \\ \hline
\rowcolor[HTML]{EBE9F5} 
SPAI\_PIG  & 187 & 2357016 & 0.671 \\ \hline
\rowcolor[HTML]{EBE9F5} 
WAP3\_PIG & 144 & 1649378 & 0.610 \\ \hline
\end{tabular}}
\caption{Gap penalty $= 1$}
\label{QSWP1}
\end{subtable}
\hfill
\begin{subtable}{0.49\textwidth}
\centering
\resizebox{\textwidth}{!}{
\begin{tabular}{|c|c|c|c|}
\hline
\rowcolor[HTML]{A490DD} 
Protein name & \begin{tabular}[c]{@{}c@{}}Sequence \\ length\end{tabular} & Score & Similarity \\ \hline
\rowcolor[HTML]{EBE9F5} 
ELAF\_HUMAN  & 117 & 1893437 & 1.000 \\ \hline
\rowcolor[HTML]{EBE9F5} 
ELAF\_PIG  & 167 & 1559007 & 0.577 \\ \hline
\rowcolor[HTML]{EBE9F5} 
SPAI\_PIG & 187 & 1617678 & 0.535 \\ \hline
\rowcolor[HTML]{EBE9F5} 
WAP3\_PIG & 144 & 1156066 & 0.496 \\ \hline
\end{tabular}}
\caption{Gap penalty $= 2$}
\label{QSWP2}
\end{subtable}
\begin{subtable}{0.49\textwidth}
\centering
\resizebox{\textwidth}{!}{
\begin{tabular}{|c|c|c|c|}
\hline
\rowcolor[HTML]{A490DD} 
Protein name & \begin{tabular}[c]{@{}c@{}}Sequence \\ length\end{tabular} & Score & Similarity \\ \hline
\rowcolor[HTML]{EBE9F5} 
ELAF\_HUMAN  & 117 & 1487539 & 1.000 \\ \hline
\rowcolor[HTML]{EBE9F5} 
SPAI\_PIG  & 187 & 1037783 & 0.436 \\ \hline
\rowcolor[HTML]{EBE9F5} 
ELAF\_PIG  & 167 & 909223 & 0.428 \\ \hline
\rowcolor[HTML]{EBE9F5} 
WAP3\_PIG  & 144 & 691554 & 0.378 \\ \hline
\end{tabular}}
\caption{Gap penalty $= 4$}
\label{QSWP4}
\end{subtable}
\caption{Podium for Quantum Smith-Waterman with various gap penalties}
\label{QSWpodium}
\end{table}

For the human elafin tests (Table \ref{Diffpodium}), the conflict graph gives good results in terms of the podiums order and difference. For the classical algorithms, they gave the same results. However, for the quantum algorithms, that of Needleman-Wunsch (global alignment) gave better results than Smith-Waterman (local alignment). The two algorithms with results closest to the reference are: Conflict graph and Quantum Needleman-Wunsch.

\begin{table}[h!]
\centering
\begin{tabular}{|c|c|c|c|}
\hline
\rowcolor[HTML]{A490DD} 
Algorithms                   & ELAF\_PIG & WAP3\_PIG & SPAI\_PIG \\ \hline
\rowcolor[HTML]{EBE9F5} 
Conflict graph               & 3         & 1.5       & 1.8       \\ \hline
\rowcolor[HTML]{EBE9F5} 
Classical   Needleman-Wunsch & 8.4       & 1.2       & 2.9       \\ \hline
\rowcolor[HTML]{EBE9F5} 
Quantum Needleman-Wunsch     & 2.2       & 2.6       & 3.5       \\ \hline
\rowcolor[HTML]{EBE9F5} 
Classical   Smith-Waterman   & 8.4       & 1.2       & 2.9       \\ \hline
\rowcolor[HTML]{EBE9F5} 
Quantum Smith-Waterman       & 5.1       & 2.2       & 38.5      \\ \hline
\end{tabular}
\caption{Summary of the percentage difference with the BLASTP reference}
\label{Diffpodium}
\end{table}

\subsection{Results for Quantum generator}

We have generate a sequence of 50 amino acids with our quantum generator, the sequence is as follows:

\begin{equation}
\mbox{MASTIPMGMNRQTAVYFTMKGMLADASARAVQPRSVHPAHPSTHFNGTSH}
\end{equation}

BLASTP with this sequence find only one protein with a score above 80 (Table \ref{Blastp_generator}).

\begin{table}[h!]
\centering
\begin{tabular}{|c|c|c|c|c|c|c|}
\hline
\rowcolor[HTML]{A490DD} 
Protein name & \begin{tabular}[c]{@{}c@{}}Sequence \\ length\end{tabular} & \begin{tabular}[c]{@{}c@{}}Query \\ cover\end{tabular} & Score & Identity\\ \hline
\rowcolor[HTML]{EBE9F5} 
070R\_FRG3G
 & 124 & 0.94 & 90.1 & 0.94 \\ \hline
\end{tabular}
\caption{BLASP result for the generated protein}
\label{Blastp_generator}
\end{table}

We're going to test our algorithms with this sequence and see if we find the protein given by the reference, BLASTP. For the Conflict graph algorithm, the parameters taken are the same defined in the last part, for the tests with elafin. We have $\left( W_{sim}, Min_{sim}, W_{edges} , \delta \right) = \left( 0.30, 0.75,1.00,0.25 \right)$. For the others algorithms, the gap penalty is taken with a value of 4, since the latter gave the best results in tests with human elafin.

\begin{table}[h!]
\centering
\begin{subtable}{0.49\textwidth}
\centering
\resizebox{\textwidth}{!}{
\begin{tabular}{|c|c|c|c|c|}
\hline
\rowcolor[HTML]{A490DD} 
Protein   name                   & \begin{tabular}[c]{@{}c@{}}Sequence \\ length\end{tabular} & Score & Norm  & {\color[HTML]{000000} Similarity} \\ \hline
\rowcolor[HTML]{EBE9F5} 
P14\_BPPH6           & 62              & 0.567 & 0.806 & 0.457                             \\ \hline
\rowcolor[HTML]{EBE9F5} 
NDBT\_LYCMC  & 59              & 0.513 & 0.847 & 0.435                             \\ \hline
\rowcolor[HTML]{EBE9F5} 
LYS0\_ECOLX      & 49              & 0.417 & 0.980 & 0.409                             \\ \hline
\end{tabular}}
\caption{Classic Smith-Waterman}
\label{CSW_generator}
\end{subtable}
\hfill
\begin{subtable}{0.49\textwidth}
\centering
\resizebox{\textwidth}{!}{
\begin{tabular}{|c|c|c|c|c|}
\hline
\rowcolor[HTML]{A490DD} 
Protein   name                              & \begin{tabular}[c]{@{}c@{}}Sequence \\ length\end{tabular} & Score  & Norm  & {\color[HTML]{000000} Similarity} \\ \hline
\rowcolor[HTML]{EBE9F5} 
 070R\_FRG3G   & 124             & 239352 & 0.403 & 0.613                             \\ \hline
\rowcolor[HTML]{EBE9F5} 
 VG50\_ICHVA  & 670             & 755935 & 0.075 & 0.358                             \\ \hline
\rowcolor[HTML]{EBE9F5} 
DERPC\_RAT  & 533 & 571627 & 0.094 & 0.340 \\ \hline
\end{tabular}}
\caption{Quantum Smith-Waterman}
\label{QSW_generator}
\end{subtable}

\begin{subtable}{0.49\textwidth}
\centering
\resizebox{\textwidth}{!}{
\begin{tabular}{|c|c|c|c|c|}
\hline
\rowcolor[HTML]{A490DD} 
Protein   name             & \begin{tabular}[c]{@{}c@{}}Sequence \\ length\end{tabular} & Score & Norm  & {\color[HTML]{000000} Similarity} \\ \hline
\rowcolor[HTML]{EBE9F5} 
RL333\_STRCO & 54              & 0.421 & 0.926 & 0.390                             \\ \hline
\rowcolor[HTML]{EBE9F5} 
RL32\_PARL1  & 60              & 0.445 & 0.833 & 0.371                             \\ \hline
\rowcolor[HTML]{EBE9F5} 
 RL331\_MYCBO & 54              & 0.393 & 0.926 & 0.364                             \\ \hline
\end{tabular}}
\caption{Classical Needleman-Wunsch}
\label{CNM_generator}
\end{subtable}
\hfill
\begin{subtable}{0.49\textwidth}
\centering
\resizebox{\textwidth}{!}{
\begin{tabular}{|c|c|c|c|c|}
\hline
\rowcolor[HTML]{A490DD} 
Protein   name                                  & \begin{tabular}[c]{@{}c@{}}Sequence \\ length\end{tabular} & Score & Norm  & {\color[HTML]{000000} Similarity} \\ \hline
\rowcolor[HTML]{EBE9F5} 
P14\_BPPH6                          & 62              & 46314 & 0.806 & 0.2475                            \\ \hline
\rowcolor[HTML]{EBE9F5} 
RL37\_METB6         & 60              & 43978 & 0.833 & 0.243                             \\ \hline
\rowcolor[HTML]{EBE9F5} 
QCR9\_MOUSE  & 64              & 43853 & 0.781 & 0.227                             \\ \hline
\end{tabular}}
\caption{Quantum Needleman-Wunsch}
\label{QNM_generator}
\end{subtable}
\begin{subtable}{0.49\textwidth}
\centering
\resizebox{\textwidth}{!}{
\begin{tabular}{|c|c|c|c|c|}
\hline
\rowcolor[HTML]{A490DD} 
Protein   name                                  & \begin{tabular}[c]{@{}c@{}}Sequence \\ length\end{tabular} & Score & Norm  & {\color[HTML]{000000} Similarity} \\ \hline
\rowcolor[HTML]{EBE9F5} 
 070R\_FRG3G    & 124             & 0.656 & 0.403 & 0.264                             \\ \hline
\rowcolor[HTML]{EBE9F5} 
Y5538\_DICDI  & 72              & 0.304 & 0.694 & 0.211                             \\ \hline
\rowcolor[HTML]{EBE9F5} 
RL34\_PSEF5        & 44              & 0.175 & 0.880 & 0.154                             \\ \hline
\end{tabular}}
\caption{Conflict graph}
\label{CG_generator}
\end{subtable}
\caption{Podiums of the different algorithms for the generated protein}
\label{Generator_podium}
\end{table}

Only two algorithms give the same protein than BLASTP in the first place,  Conflict graph and Quantum Smith-Waterman algorithms.

In view of the above, we will use the Conflict graph algorithm or the Smith-Waterman algorithm to compare proteins. The Needleman-Wunsch algorithm does not find the protein for the generator, even though it gave good results for tests with elafin. On the other hand, the scoring matrix is less suitable for QAOA. On the other hand, when tested with the generator protein, the Smith-Waterman quantum algorithm gives the same results as the reference, unlike the classical version, so the quantum version is more robust.

%========================================================================================
\section{Discussion and conclusions}
%========================================================================================

First, a random amino-acid's sequence generator was made for a quantum formalism and respecting certain conditions given by the database. The advantage of this generator is that you can create sequences by taking into account the appearance of the different amino acids and their positions (probability law). Moreover, the choice of amino acids is made by measuring quantum states, which ensures randomness. The process described for the quantum sequence generator is used to initialise a quantum system into a given state.  

With this in mind, we have constructed a process for comparing proteins using two different methods: the first using the matrix of two bioinformatics algorithms (Needleman-Wunsch and Smith-Waterman), and the second based on the adjacency matrix of a graph that will be constructed in relation to the properties of the two proteins being compared. All the algorithms have been either extended with a quantum formalism (Quantum Needleman-Wunsch and Smith-Waterman algorithm) or extended for the case of amino-acids sequence (Conflict graph algorithm).

 The main result we obtained from numerical simulations is that for the parameters defined in this article ($W_{sim} = 0.30, Min_{sim} = 0.75, W_{edges} = 1.00 , \delta = 0.25$), the Conflict graph algorithm gives the same result than our reference both in terms of the order of the podium and the difference with BLASTP, like Quantum Needleman-Wunsch Algorithm with a gap penalty value of $4$. However, Quantum Needleman-Wunsch does not give good results for a random protein provided with the quantum generator contrary to Quantum Smith-Waterman algorithm. Moreover, the structure of the scoring matrix constructed with the Needleman-Wunsch algorithm is less adapted for QAOA. QAOA has difficulty finding the fundamental state of the quantum system due the spectrum of the scoring matrix is globally symmetrical, resulting in a very slow resolution time, which is not the case for the Smith-Waterman or the Conflict graph algorithms.

In light of some encouraging results on the quantum contribution for drug discovery, it would be interesting to make studies on other drug discovery problematics and implement quantum algorithm. In the research and development phase of the pharmaceutical value chain, two interesting research topics are,  the transition from hit to lead and the improvement of the lead. But also to improve the Conflict graph algorithm by taking into account the fingerprint of proteins, for example or explore in more detail the Needleman-Wunsch algorithm and see why the convergence of QAOA is slower and validate or not the hypothesis made in this article, as this is due to the near symmetry of the spectrum of the matrix.

\section*{Acknowledgements}

This work was made possible thanks to funding from the Expleo Group for its in-house R\&I project. This work is included in the project Deep Microbiota supported by Expleo Group. The views and opinions of the authors expressed herein do not necessarily state or reflect those of Expleo Group or any collaborators thereof.

I'd like to thank Julien Mellaerts very much for his program, which served as the basis for our Conflict graph algorithm program. I would also like to thank the people who contributed to Expleo's Deep Microbiota project in their own way: Sonia Saidani, Rahma Lemoudda, Meriem Benhabib and Delphine Sannier.

%========================================================================================
\section*{Conflict of interest}
The authors declare that they have no known competing financial interests or personal relationships that could have appeared to influence the work reported in this paper.
%========================================================================================

\bibliographystyle{quantum}
\bibliography{refs}

\end{document}